\documentclass[twocolumn]{aastex631}

\usepackage{amsmath}
\usepackage{float}
\usepackage{graphicx}
\usepackage{gensymb}  
\usepackage{multirow}  

\submitjournal{ApJ}

\shorttitle{JWST High-resolution Spectroscopy of \JD}
\shortauthors{Abdurro'uf et al.}

\begin{document}

\title{JWST NIRSpec High-resolution Spectroscopy of \JD\ at $z=10.167$:\\ Resolved [OII] Doublet and Electron Density in an Early Galaxy}

\newcommand{\HubbleFellow}{\altaffiliation{Hubble Fellow}}

\correspondingauthor{Abdurro'uf}
\email{fabdurr1@jhu.edu}

\author[0000-0002-5258-8761]{Abdurro'uf}
\affiliation{Department of Physics and Astronomy, The Johns Hopkins University, 3400 N Charles St. Baltimore, MD 21218, USA}
\affiliation{Space Telescope Science Institute (STScI), 3700 San Martin Drive, Baltimore, MD 21218, USA}

\author[0000-0003-2366-8858]{Rebecca L. Larson}
\affiliation{School of Physics and Astronomy, Rochester Institute of Technology, 84 Lomb Memorial Drive, Rochester, NY 14623, USA}

\author[0000-0001-7410-7669]{Dan Coe}
\affiliation{Space Telescope Science Institute (STScI), 3700 San Martin Drive, Baltimore, MD 21218, USA}
\affiliation{Association of Universities for Research in Astronomy (AURA), Inc.~for the European Space Agency (ESA)}
\affiliation{Department of Physics and Astronomy, The Johns Hopkins University, 3400 N Charles St. Baltimore, MD 21218, USA}

\author[0000-0003-4512-8705]{Tiger Yu-Yang Hsiao}
\affiliation{Department of Physics and Astronomy, The Johns Hopkins University, 3400 N Charles St. Baltimore, MD 21218, USA}
\affiliation{Space Telescope Science Institute (STScI), 3700 San Martin Drive, Baltimore, MD 21218, USA}

\author[0000-0002-7093-1877]{Javier Álvarez-Márquez}
\affiliation{Centro de Astrobiología (CAB), CSIC-INTA, Ctra. de Ajalvir km 4, Torrejón de Ardoz, E-28850, Madrid, Spain}

\author[0000-0003-2119-277X]{Alejandro Crespo Gómez}
\affiliation{Centro de Astrobiología (CAB), CSIC-INTA, Ctra. de Ajalvir km 4, Torrejón de Ardoz, E-28850, Madrid, Spain}

\author[0000-0002-8192-8091]{Angela Adamo}
\affiliation{Department of Astronomy, The Oskar Klein Centre, Stockholm University, AlbaNova, SE-10691 Stockholm, Sweden}

\author[0000-0003-0883-2226]{Rachana Bhatawdekar}
\affiliation{European Space Agency (ESA), European Space Astronomy Centre (ESAC), Camino Bajo del Castillo s/n, 28692 Villanueva de la Cañada, Madrid, Spain}

\author[0000-0001-8068-0891]{Arjan Bik}
\affiliation{Department of Astronomy, Stockholm University, Oscar Klein Centre, AlbaNova University Centre, 106 91 Stockholm, Sweden}

\author[0000-0002-7908-9284]{Larry D. Bradley}
\affiliation{Space Telescope Science Institute (STScI), 3700 San Martin Drive, Baltimore, MD 21218, USA}

\author[0000-0003-1949-7638]{Christopher J. Conselice}
\affiliation{Jodrell Bank Centre for Astrophysics, University of Manchester, Oxford Road, Manchester UK}

\author[0000-0001-8460-1564]{Pratika Dayal}
\affiliation{Kapteyn Astronomical Institute, University of Groningen, 9700 AV Groningen, The Netherlands}

\author[0000-0001-9065-3926]{Jose M. Diego}
\affiliation{Instituto de F\'isica de Cantabria (CSIC-UC). Avda. Los Castros s/n. 39005 Santander, Spain}

\author[0000-0001-7201-5066]{Seiji Fujimoto}\altaffiliation{Hubble Fellow}
\affiliation{The University of Texas at Austin, Department of Astronomy, Austin, TX, United States}

\author[0000-0001-6278-032X]{Lukas J. Furtak}
\affiliation{Physics Department, Ben-Gurion University of the Negev, P.O. Box 653, Be'er-Sheva 84105, Israel}

\author[0000-0003-4372-2006]{Taylor A. Hutchison}\altaffiliation{NASA Postdoctoral Fellow}
\affiliation{Observational Cosmology Lab, NASA Goddard Space Flight Center, Greenbelt, MD 20771, USA}

\author[0000-0003-1187-4240]{Intae Jung}
\affil{Space Telescope Science Institute, 3700 San Martin Drive Baltimore, MD 21218, United States}

\author[0000-0001-5289-3291]{Meghana Killi}
\affiliation{Instituto de Estudios Astrofísicos, Facultad de Ingeniería y Ciencias, Universidad Diego Portales, Av. Ejército 441, Santiago 8370191, Chile}

\author[0000-0002-5588-9156]{Vasily Kokorev}
\affiliation{Kapteyn Astronomical Institute, University of Groningen, P.O. Box 800, 9700 AV Groningen, The Netherlands}

\author[0000-0003-2589-762X]{Matilde Mingozzi}
\affiliation{Space Telescope Science Institute (STScI), 3700 San Martin Drive, Baltimore, MD 21218, USA}

\author[0000-0002-5222-5717]{Colin Norman}
\affiliation{Department of Physics and Astronomy, The Johns Hopkins University, 3400 N Charles St. Baltimore, MD 21218, USA}
\affiliation{Space Telescope Science Institute (STScI), 3700 San Martin Drive, Baltimore, MD 21218, USA}

\author[0009-0007-0522-7326]{Tom Resseguier}
\affiliation{Department of Physics and Astronomy, The Johns Hopkins University, 3400 N Charles St. Baltimore, MD 21218, USA}
\affiliation{Space Telescope Science Institute (STScI), 3700 San Martin Drive, Baltimore, MD 21218, USA}

\author[0000-0003-4223-7324]{Massimo Ricotti}
\affiliation{Department of Astronomy, University of Maryland, College Park, 20742, USA}

\author[0000-0002-7627-6551]{Jane R.~Rigby}
\affiliation{Observational Cosmology Lab, NASA Goddard Space Flight Center, Greenbelt, MD 20771, USA}

\author[0000-0002-5057-135X]{Eros Vanzella}
\affiliation{INAF -- OAS, Osservatorio di Astrofisica e Scienza dello Spazio di Bologna, via Gobetti 93/3, I-40129 Bologna, Italy}

\author[0000-0003-1815-0114]{Brian Welch}
\affiliation{Observational Cosmology Lab, NASA Goddard Space Flight Center, Greenbelt, MD 20771, USA}
\affiliation{Department of Astronomy, University of Maryland, College Park, 20742, USA}

\author[0000-0001-8156-6281]{Rogier A. Windhorst} 
\affiliation{School of Earth and Space Exploration, Arizona State University, Tempe, AZ 85287-1404, USA}

\author[0000-0002-9217-7051]{Xinfeng Xu}
\affiliation{Department of Physics and Astronomy, Northwestern University, 2145 Sheridan Road, Evanston, IL, 60208, USA}

\author[0000-0002-0350-4488]{Adi Zitrin}
\affiliation{Physics Department, Ben-Gurion University of the Negev, P.O. Box 653, Be'er-Sheva 84105, Israel}





\newcommand{\LCDM}{$\Lambda$CDM}

\newcommand{\red}[1]{{\color{red} #1}}
\newcommand{\redss}[1]{{\color{red} ** #1}}
\newcommand{\redbf}[1]{{\color{red}\bf #1 \color{black}}}

\newcommand{\ny}{$\tilde {\rm n}$}
\newcommand{\about}{$\sim$}
\newcommand{\appr}{$\approx$}
\newcommand{\gt}{$>$}
\newcommand{\uJy}{$\mu$Jy}
\newcommand{\sig}{$\sigma$}
\newcommand{\Lya}{Lyman-$\alpha$}
\renewcommand{\th}{$^{\rm th}$}
\newcommand{\lam}{$\lambda$}

\newcommand{\tentothe}[1]{$10^{#1}$}
\newcommand{\tentotheminus}[1]{$10^{-#1}$}
\newcommand{\e}[1]{$\times 10^{#1}$}
\newcommand{\en}[1]{$\times 10^{-#1}$}
\newcommand{\cgsfluxunits}{erg$\,$s$^{-1}\,$cm$^{-2}$}
\newcommand{\cgsfluxdensityunits}{erg$\,$s$^{-1}\,$cm$^{-2}$\,\AA$^{-1}$}
\newcommand{\linefluxunits}{\tentotheminus{20} \cgsfluxunits}

\newcommand{\logU}{$\log(U)$}
\newcommand{\logOH}{12+log(O/H)}

\newcommand{\sinv}{s$^{-1}$}

\newcommand{\footnoteurl}[1]{\footnote{\url{#1}}}

\newcommand{\tnm}[1]{\tablenotemark{#1}}
\newcommand{\super}[1]{$^{\rm #1}$}
\newcommand{\supa}{$^{\rm a}$}
\newcommand{\supb}{$^{\rm b}$}
\newcommand{\supc}{$^{\rm c}$}
\newcommand{\supd}{$^{\rm d}$}
\newcommand{\supe}{$^{\rm e}$}
\newcommand{\supf}{$^{\rm f}$}
\newcommand{\supg}{$^{\rm g}$}
\newcommand{\suph}{$^{\rm h}$}
\newcommand{\supi}{$^{\rm i}$}
\newcommand{\supj}{$^{\rm j}$}
\newcommand{\supk}{$^{\rm k}$}
\newcommand{\supl}{$^{\rm l}$}
\newcommand{\supm}{$^{\rm m}$}
\newcommand{\supn}{$^{\rm n}$}
\newcommand{\supo}{$^{\rm o}$}

\newcommand{\sqarcmin}{arcmin\squared}

\newcommand{\supcomma}{$^{\rm ,}$}

\newcommand{\rhalf}{$r_{1/2}$}

\newcommand{\chisq}{$\chi^2$}

\newcommand{\Zgas}{$Z_{\rm gas}$}  
\newcommand{\Zstar}{$Z_*$}  

\newcommand{\inv}{\per}
\newcommand{\Mstar}{$M^*$}
\newcommand{\Lstar}{$L^*$}
\newcommand{\phistar}{$\phi^*$}

\newcommand{\logM}{log($M_*$/\Msun)}

\newcommand{\LUV}{$L_{UV}$}
\newcommand{\MUV}{$M_{UV}$}

\newcommand{\Msun}{$M_\odot$}
\newcommand{\Lsun}{$L_\odot$}
\newcommand{\Zsun}{$Z_\odot$}

\newcommand{\Mvir}{$M_{vir}$}
\newcommand{\Mt}{$M_{200}$}
\newcommand{\Mf}{$M_{500}$}

\newcommand{\Ndotion}{$\dot{N}_{\rm ion}$}
\newcommand{\xiion}{$\xi_{\rm ion}$}
\newcommand{\logxiion}{log(\xiion)}
\newcommand{\fesc}{$f_{\rm esc}$}

\newcommand{\XHI}{$X_{\rm HI}$}
\newcommand{\XHII}{$X_{\rm HII}$}
\newcommand{\RHII}{$R_{\rm HII}$}

\newcommand{\Halpha}{H$\alpha$}
\newcommand{\Hbeta}{H$\beta$}
\newcommand{\Hgamma}{H$\gamma$}
\newcommand{\Hdelta}{H$\delta$}
\newcommand{\Halphaw}{\Halpha\,$\lambda$6563}
\newcommand{\Hbetaw}{\Hbeta\,$\lambda$4861}
\newcommand{\Hgammaw}{H$\gamma$\,$\lambda$4342}
\newcommand{\Hdeltaw}{H$\delta$\,$\lambda$4103}
\newcommand{\Ha}{\Halpha}
\newcommand{\Hb}{\Hbeta}

\newcommand{\I}{\,{\sc i}}
\newcommand{\II}{\,{\sc ii}}
\newcommand{\III}{\,{\sc iii}}
\newcommand{\IV}{\,{\sc iv}}
\newcommand{\VI}{\,{\sc vi}}
\newcommand{\VII}{\,{\sc vii}}
\newcommand{\VIII}{\,{\sc viii}}

\newcommand{\HI}{H\,{\sc i}}
\newcommand{\HII}{H\,{\sc ii}}
\newcommand{\HeI}{He\,{\sc i}}
\newcommand{\HeII}{He\,{\sc ii}}

\newcommand{\CII}{[C\,{\sc ii}]}
\newcommand{\CIIw}{\CII\,$\lambda$2325 (blend)}
\newcommand{\CIII}{C\,{\sc iii}]}
\newcommand{\CIIIwa}{\CIII\,$\lambda$1907}
\newcommand{\CIIIwb}{\CIII\,$\lambda$1909}
\newcommand{\CIIId}{C\,{\sc iii}]}
\newcommand{\CIIIdw}{C\,{\sc iii}]\,$\lambda\lambda$1907,1909}
\newcommand{\CIV}{C\,{\sc iv}}
\newcommand{\CIVw}{\CIV\,$\lambda$1549}
\newcommand{\OII}{[O\,{\sc ii}]}
\newcommand{\OIIwa}{\OII\,$\lambda$3726}
\newcommand{\OIIwb}{\OII\,$\lambda$3729}
\newcommand{\OIIdw}{\OII\,$\lambda\lambda$3726,3729}
\newcommand{\OIII}{[O\,{\sc iii}]}
\newcommand{\OIIIw}{\OIII\,$\lambda$5007}
\newcommand{\OIIIww}{\OIII\,$\lambda\lambda$4959,5007}
\newcommand{\OIIIwa}{\OIII\,$\lambda$4363}
\newcommand{\OIIIwc}{\OIII\,$\lambda$4959}
\newcommand{\NeIII}{[Ne\,{\sc iii}]}
\newcommand{\NeIIIw}{\NeIII\,$\lambda$3870}
\newcommand{\NeIIIwb}{\NeIII\,$\lambda$3969}
\newcommand{\HeIw}{HeI\,$\lambda$3890}
\newcommand{\HeIwa}{HeI\,$\lambda$4473}
\newcommand{\HeIIw}{HeII\,$\lambda$1640}
\newcommand{\NII}{[N\,{\sc ii}]}
\newcommand{\NIII}{N\,{\sc iii}]}
\newcommand{\NIV}{N\,{\sc iv}]}
\newcommand{\NIIIw}{\NIII\,$\lambda$1748}
\newcommand{\NIVw}{\NIV\,$\lambda$1486}
\newcommand{\MgII}{Mg\,{\sc ii}}
\newcommand{\MgIIw}{\MgII\,$\lambda$2800}

\newcommand{\SII}{[S\,{\sc ii}]}
\newcommand{\SIIwa}{\SII\,$\lambda$6716}
\newcommand{\SIIwb}{\SII\,$\lambda$6731}
\newcommand{\SIIdw}{\SII\,$\lambda\lambda$6716,6731}

\newcommand{\Lyaw}{Ly$\alpha$\,$\lambda$1216}

\newcommand{\OIIIwfivem}{\OIII\,$\lambda$52$\mu$m}
\newcommand{\OIIIweightm}{\OIII\,$\lambda$88$\mu$m}



\newcommand{\Om}{\Omega_{\rm M}}
\newcommand{\OL}{\Omega_\Lambda}

\newcommand{\etal}{et al.}

\newcommand{\citeps}{\citep}

\newcommand{\HST}{{\em HST}}
\newcommand{\SST}{{\em SST}}
\newcommand{\Hubble}{{\em Hubble}}
\newcommand{\Spitzer}{{\em Spitzer}}
\newcommand{\Chandra}{{\em Chandra}}
\newcommand{\JWST}{{\em JWST}}
\newcommand{\Planck}{{\em Planck}}

\newcommand{\Bradac}{{Brada\v{c}}}

\newcommand{\citepeg}[1]{\citep[e.g.,][]{#1}}

\newcommand{\range}[2]{\! \left[ _{#1} ^{#2} \right] \!}  

\newcommand{\grizli}{\textsc{grizli}}
\newcommand{\eazypy}{\textsc{eazypy}}
\newcommand{\msaexp}{\textsc{msaexp}}
\newcommand{\trilogy}{\textsc{trilogy}}
\newcommand{\bagpipes}{\textsc{bagpipes}}
\newcommand{\beagle}{\textsc{beagle}}
\newcommand{\photutils}{\textsc{photutils}}
\newcommand{\SEP}{\textsc{sep}}
\newcommand{\piXedfit}{\textsc{piXedfit}}
\newcommand{\pyneb}{\textsc{pyneb}}
\newcommand{\HIIC}{\textsc{hii-chi-mistry}}
\newcommand{\astropy}{\textsc{astropy}}
\newcommand{\astrodrizzle}{\textsc{astrodrizzle}}
\newcommand{\multinest}{\textsc{multinest}}
\newcommand{\cloudy}{\textsc{Cloudy}}
\newcommand{\jdaviz}{\textsc{Jdaviz}}
\newcommand{\emcee}{\textsc{emcee}}

\renewcommand{\tt}[1]{\texttt{#1}}

\newcommand{\SE}{\tt{SourceExtractor}}

\newcommand{\PD}[1]{\textcolor{blue}{[PD: #1\;]}}

\newcommand{\JD}{MACS0647$-$JD}

\newcommand{\edense}{$n_{e}$}
\newcommand{\ROII}{$R_{[\rm{OII}]}$}
\newcommand{\OIIratio}{\OII\,$\lambda$3729/$\lambda$3726}

\newcommand{\RSII}{$R_{[\rm{SII}]}$}
\newcommand{\SIIratio}{\SII\,$\lambda$6716/$\lambda$6731}
\newcommand{\CIIIratio}{C\,{\sc iii}]\,$\lambda$1909/$\lambda$1907}

\newcommand{\lya}{\hbox{Ly$\alpha$}}        
\newcommand{\nv}{\hbox{\sc N\,v}}           
\newcommand{\niv}{\hbox{\sc N\,iv]}}      
\newcommand{\ariv}{\hbox{\sc Ar\,iv]}}      
\newcommand{\civ}{\hbox{\sc C\,iv}}         
\newcommand{\heii}{\hbox{He\,{\sc ii}}}     
\newcommand{\cii}{\hbox{\sc C\,ii]}}      
\newcommand{\hei}{\hbox{He\,{\sc i}}}     
\newcommand{\oiiisemi}{\hbox{\sc O\,iii]}}  
\newcommand{\ciii}{\hbox{\sc C\,iii]}}      
\newcommand{\siiii}{\hbox{Si\,{\sc iii]}}}  
\newcommand{\mgii}{\hbox{Mg\,{\sc ii}}}     
\newcommand{\oi}{\hbox{O\,{\sc i}}}     
\newcommand{\oii}{\hbox{\sc [O\,ii]}}     
\newcommand{\hb}{\hbox{\sc H$\beta$}}       
\newcommand{\oiii}{\hbox{\sc [O\,iii]}}     
\newcommand{\ha}{\hbox{\sc H$\alpha$}}      
\newcommand{\hd}{\hbox{\sc H$\delta$}}      
\newcommand{\hg}{\hbox{\sc H$\gamma$}}      
\newcommand{\he}{\hbox{\sc H$\epsilon$}}      
\newcommand{\het}{\hbox{\sc H$\eta$}}      
\newcommand{\sii}{\hbox{[S\,{\sc ii}]}}     
\newcommand{\siii}{\hbox{[S\,{\sc iii}]}}   
\newcommand{\pab}{\hbox{Pa$\beta$}}      
\newcommand{\pag}{\hbox{Pa$\gamma$}}
\newcommand{\neiii}{\hbox{[Ne\,{\sc iii]}}}  
\newcommand{\siiiIR}{\hbox{[S\,{\sc iii]}}}  
\newcommand{\feii}{\hbox{[Fe\,{\sc ii]}}}  
\newcommand{\nev}{\hbox{[Ne\,{\sc v]}}}  
\newcommand{\oiv}{\hbox{O\,{\sc iv]}}}  
\newcommand{\siv}{\hbox{[S\,{\sc iv]}}}  

\newcommand{\kms}{km\,s$^{-1}$}

\begin{abstract}
We present JWST/NIRSpec high-resolution spectroscopy G395H/F290LP of \JD, a gravitationally lensed galaxy merger at $z=10.167$. The new spectroscopy, which is acquired for the two lensed images (JD1 and JD2), detects and resolves emission lines in the rest-frame ultraviolet (UV) and blue optical, including the resolved \OIIdw\ doublet, \NeIIIw, \HeIw, \Hdelta, \Hgamma, and \OIIIwa. This is the first observation of the resolved \OIIdw\ doublet for a galaxy at $z>8$. We measure a line flux ratio \OIIratio\ $= 0.9 \pm 0.3$, which corresponds to an estimated electron density of $\log(n_{e} / \rm{cm}^{-3}) = 2.9 \pm 0.5$. 
This is significantly higher than the electron densities of local galaxies reported in the literature. We compile the measurements from the literature and further analyze the redshift evolution of $n_{e}$. We find that the redshift evolution follows the power-law form of $n_{e} = A\times (1+z)^{p}$ with $A=54^{+31}_{-23}$ cm$^{-3}$ and $p=1.2^{+0.4}_{-0.4}$. This power-law form may be explained by a combination of metallicity and morphological evolution of galaxies, which become, on average, more metal-poor and more compact with increasing redshift.
\end{abstract}


\section{Introduction} \label{sec:intro}
 
Understanding the physical conditions in star-forming galaxies, including the interstellar medium (ISM), is essential for a complete picture of the evolution of the stellar and gaseous content of galaxies. Spectroscopy of galaxies in the rest-frame ultraviolet (UV) and optical provides a wealth of emission line diagnostics for characterizing the ISM in galaxies \citep[e.g., see review by][]{Kewley2019}. The relative strengths of the emission lines are mainly driven by the ISM properties, including the chemical abundance, shape of the ionizing radiation field, ionization state, gas density, and temperature \citep[e.g.,][]{Kewley2002,Dopita2006a,Dopita2006b}.

Before the launch of the JWST, the observations of rest-frame UV and optical spectra of high-redshift ($z \gtrsim 3$) galaxies were very challenging because of the lack of highly sensitive (i.e.,~deep) near-infrared (NIR) spectroscopy. Only a few spectroscopic surveys have been conducted to observe rest-frame UV and optical spectroscopy of high-$z$ galaxies out to $z\sim 4$ using 8--10 m class telescopes \citep[e.g.,][]{Kriek2015,Lefevre2015,Bacon2017,Pentericci2018,Urrutia2019}. 

Since the launch of JWST \citep{Rigby2023,Gardner2023}, with its NIR and mid-infrared (MIR) spectroscopic capabilities, the rest-frame UV and optical emission lines of $z>2$ galaxies have become easily accessible, enabling the study of the ISM properties in galaxies at high redshift (toward the reionization era). Several studies use JWST/Near Infrared Spectrograph (NIRSpec) spectroscopy \citep{Jakobsen2022,Boker2023} to study the properties of ISM in galaxies up to $z\sim 9$ \citep[e.g.,][]{Fujimoto2022, Curti2023, Fujimoto2023, Tang2023, Bunker2023, Heintz2023a, Heintz2023b, Reddy2023, Nakajima2023, Hsiao2023b, Isobe2023, Jung2023, Cameron2023, Sanders2024, Backhaus2024}. Those studies found that high-$z$ galaxies have significantly lower gas-phase metallicity \citep[e.g.,][]{Fujimoto2023, Heintz2023a, Nakajima2023, Williams2023} and higher electron density (\edense; e.g.,~\citealt{Reddy2023,Isobe2023}) than low-$z$ galaxies.      

The electron density of the \HII\ regions is one of the important properties of the ISM because together with ISM pressure, they govern the emission from \HII\ regions, such that the derived quantities from the nebular emission lines depend critically on assumptions about \edense\ and pressure of the nebula \citep{Kewley2019_a}.  
This quantity can be estimated using density-sensitive emission line ratios, including 
\OIIratio, \SIIratio, \CIIIratio\ \citep[e.g.,][]{Kewley2019}. In star-forming regions, the excitation energy between the two lines of those doublets is proportional to the thermal electron density. Therefore, the relative excitation rates depend only on the collision strengths of the electrons \citep{Osterbrock1989}. This makes the flux ratio between the two lines in the doublets a perfect diagnostic for the electron density. While they are all density tracers, the derived \edense\ from those lines are not necessarily comparable with each other because they are tracing different ionization regions. For instance, \OIIratio\ and \SIIratio\ trace low-ionization regions, while \CIIIratio\ traces intermediate-ionization regions \citep[e.g.,][]{Berg2021, Mingozzi2022}.  

Before JWST, several large spectroscopic surveys investigated the evolution of the electron density based on \OIIdw\ and \SIIdw\ (i.e.,~from low-ionization regions). They found that $n_{e}$ declines from $\sim 200$ cm$^{-3}$ at $z\sim 3$ to $\sim 30$ cm$^{-3}$ at $z\sim 0$ \citep[e.g.,][]{Steidel2014, Sanders2016, Kaasinen2017, Kashino2017, Harshan2020, Davies2021, Berg2022}. Due to the lack of highly sensitive NIR spectroscopy, there was a lack of $n_{e}$ studies at $z\gtrsim 3$. A few exceptions come from studies that estimated $n_{e}$ based on doubly-ionized regions, such as \citet{Killi2023} who observed a galaxy at $z=7.133$ with Atacama Large Millimeter Array (ALMA) and estimated $n_{e} \lesssim 500$ cm$^{-3}$ based on the \OIIIwfivem/\OIIIweightm\ ratio. However, this diagnostic probes a very different ISM region (i.e.,~dust-obscured) and ionization level compared to the \OIIdw\ and \SIIdw\ doublets. 

After the launch of JWST, several studies have already explored $n_{e}$ out to $z\sim 9$ using various diagnostics. \citet{Isobe2023} used the spectroscopic data from public JWST surveys to measure $n_{e}$ of 14 galaxies at $z=4.02-8.68$. Among the sample, only three galaxies (with the highest $z=7.87$) observed with NIRSpec high-resolution grating by the GLASS survey \citep{Treu2022} showed resolved \OIIdw\ doublet, while the doublet remains unresolved in the rest of the sample. They estimated $n_{e} \gtrsim 300$ cm$^{-3}$. \citet{Reddy2023} investigated the connections between the ionization parameter ($U$), electron density, and star formation rate (SFR) based on 48 galaxies at $2.7<z<6.3$. They measured $n_{e}$ based on the \SIIdw\ doublet resolved in NIRSpec medium-resolution spectra from the Cosmic Evolution Early Release Science \citep[CEERS;][]{Finkelstein2023, Fujimoto2023}. Using stacked spectra of two equal-number bins of O32$=$\OIIIww/\OIIdw, they estimated that galaxies with higher O32 have average $n_{e} \simeq 500$ cm$^{-3}$ that are $\gtrsim 5$ times larger than that of lower-O32 galaxies. The two bins have an average O32 of $1.978 \pm 0.057$ and $6.055 \pm 0.174$, while the entire sample has an average O32 of $3.457 \pm 0.074$. They also found a highly significant positive correlation between $U$ and SFR surface density ($\Sigma_{\rm SFR}$), which appears to be independent of redshift at $1.6 \lesssim z \lesssim 6.3$. 

While JWST/NIRSpec prism spectroscopy cannot resolve \OIIdw\ and \SIIdw\ lines due to low spectral resolution ($R\sim 30-300$), some studies have combined \OIIIw\ observed by this instrument with \OIIIweightm\ observed with ALMA to indirectly estimate $n_{e}$ of galaxies at $z=9.11$ \citep[$\sim 400$ cm$^{-3}$;][]{Stiavelli2023} and $z=8.496$ \citep[$\sim 220$ cm$^{-3}$;][]{Fujimoto2022} using photoionization modeling. Until now, no studies have directly measured $n_{e}$ from resolved \OIIdw\ doublet at $z>8$, while \SIIdw\ is redshifted beyond the wavelength coverage of NIRSpec. Detecting \OIIdw\ doublet remains challenging because most galaxies at such high redshift are faint. More importantly, those galaxies typically have a high O32 ratio and ionization parameter, such that more oxygen atoms are ionized to [OIII], thereby making [OII] relatively weaker, even for the same observed brightness. Moreover, high-resolution ($R>2000$) spectroscopy is needed to be able to resolve the doublets, which is only achievable using the NIRSpec high-resolution grating. 
Recently, the \OIIdw\ doublet has been observed in the spectrum of GN-z11 by \citet{Bunker2023}, one of the brightest galaxies known at $z>10$ \citep{Oesch2016,Tacchella2023}. However, the NIRSpec medium-resolution grating cannot resolve the single lines. 

In this paper, we present NIRSpec high-resolution ($R \sim 2700$) spectroscopy of \JD, a triply-lensed galaxy at $z=10.167$ \citep{Coe2013,Hsiao2023a,Hsiao2023b}. Previously, JWST/NIRCam imaging and NIRSpec prism spectroscopy have been obtained for this galaxy by GO 1433 (PI~Coe). The NIRCam imaging resolved the galaxy into two components A and B, suggesting a possible galaxy merger \citep{Hsiao2023a}. NIRSpec prism spectra of JD1 and JD2 (the two brightest lensed images of \JD) were presented in \citet{Hsiao2023b}, revealing 7 strong emission lines and becoming one of the highest redshift galaxies with detected rest-frame UV emission lines besides GHZ2/GLASS-z12 \citep{Castellano2024}, GN-z11 \citep{Bunker2023}, and  JADES-GS-z11-0 \citep{Hainline2024}. Here, we observe JD1 and JD2 with the high-resolution G395H/F290LP setting to further resolve the lines and obtain better measurements of their fluxes and widths. We resolve \OIIdw\ doublet for the first time at $z>8$ and use it to estimate $n_{e}$.      

The paper is organized as follows. We describe the NIRSpec data and its reduction in Section~\ref{sec:data}. The measurements of emission lines are described in Section~\ref{sec:analysis}. We present and discuss our results in Section~\ref{sec:result_discuss}, and finally conclude in Section~\ref{sec:summary}. Throughout, we assume the cosmological parameters of $\Omega_{m}=0.3$, $\Omega_{\Lambda}=0.7$, and $H_{0}=70\text{ km}\text{ s}^{-1}\text{ Mpc}^{-1}$.

\section{Data} \label{sec:data}

\subsection{NIRSpec MSA Spectroscopy} \label{sec:observation}
The NIRSpec micro-shutter assembly (MSA) Spectroscopy \citet{2022Ferruit} observation was conducted on January 14, 2024, as part of GO 4246 (PI~Abdurro'uf), which focuses on \JD\ at $z = 10.17$. We use the high-resolution grating G395H/F290LP that covers a wavelength range of $2.87–5.14$ $\mu$m with spectral resolution ranges from $R\sim 1900$ to $3500$ over the wavelength range. The total exposure time was 2.43 hours. Half of that time was observed with standard 3-slitlet nods. With the other half, we used single slitlets and performed two dithers. The slitlets on JD1 and JD2 mainly cover the brightest Component A (see Figure \ref{fig:macs0647jd_slit_spectra}). This is similar to the NIRSpec prism observation previously performed \citep{Hsiao2023b}. The data are all publicly available on MAST, but not the reduction we analyzed in this paper which will be described below.\footnote{\url{https://mast.stsci.edu/search/ui/\#/jwst}.}

\subsection{NIRSpec Data Reduction} \label{sec:data_reduction}

\begin{figure*}[ht]
\centering
\includegraphics[width=1.0\textwidth]{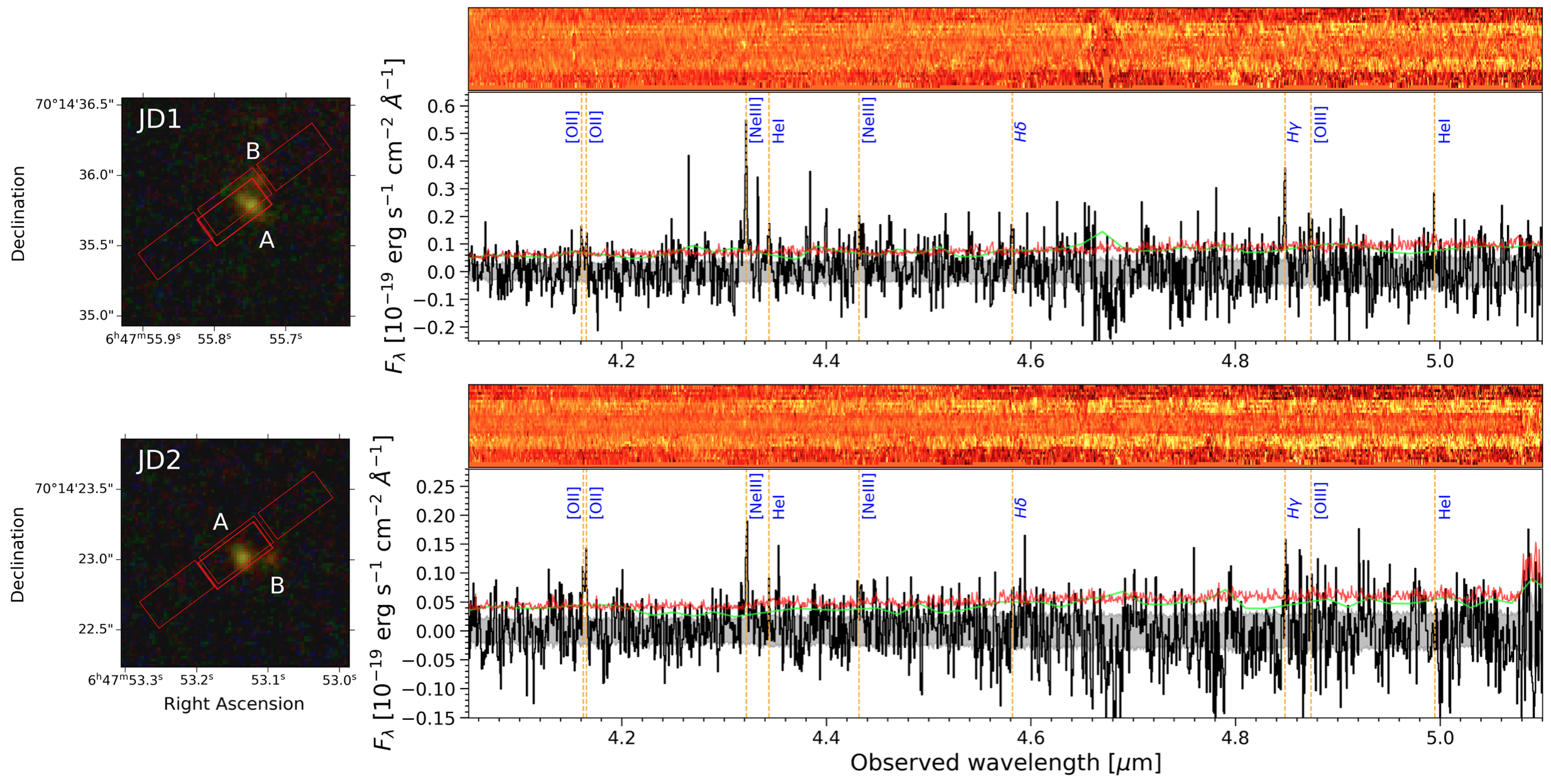}
\caption{NIRSpec G395H/F290LP spectra of \JD\ lensed images JD1 and JD2 (IDs 3593 and 3349, respectively, in the catalog used for preparing the observation). \textit{Left panel}: Slitlet positions overlaid on $1.6'' \times 1.6''$ NIRCam color images (blue:F115W; green:F150W; red:F200W). \textit{Right panel}: 2D and 1D NIRSpec high-resolution spectra of JD1 (top) and JD2 (bottom). Several emission lines are detected while the continuum is not. Emission lines are indicated with blue-colored labels and orange vertical dashed lines. The gray shaded area represents the uncertainties obtained from the pipeline. The green lines represent the standard deviations of the spectral fluxes after removing the emission lines. We multiply the flux uncertainties by a factor of two to match the overall level of the standard deviations. This is shown with a red line. These rescaled uncertainties are used for the analysis in this paper.}
\label{fig:macs0647jd_slit_spectra}
\end{figure*}

The data were reduced with the \JWST\ pipeline version 1.12.5 \citep{bushouse_2023_10022973}, using CRDS version 11.17.6 and \texttt{jwst\_1183.pmap} with the 2D spectra obtained from the Mikulski Archive for Space Telescopes (MAST). The 2D frames were visually inspected for artifacts and the snowballs and residual cosmic rays were manually masked. Additional sigma clipping with a threshold of 6$\sigma$ was then applied to the 2D spectra to remove any additional spurious pixels. The source's one-dimensional (1D) spectra were obtained via an optimal extraction \citep{Horne86} with a spatial weight profile from the trace of the source in the masked 2D spectrum such that the pixels near the peak of the trace are maximally weighted. To create the extraction profile, the 2D signal-to-noise (SNR) spectrum was collapsed in the spectral direction, taking the median value at each spectral pixel and fitting a Gaussian to the positive trace. While the NIRSpec shutter includes flux from both Component A and Component B for JD1 (ID 3593), the orientation is such that Component B falls completely out of the shutter for JD2 (ID 3349). The source is effectively spatially unresolved at all wavelengths (intrinsic size is less than the JWST point spread function full width at half maximum, PSF FWHM), and there was only a single detectable trace in both observations at the expected location of Component A. An attempt to extract a spectrum for both Component A and Component B from JD1 yielded no discernible detections for Component B, which was affected by the location of the upper negative trace in the 2D spectrum.

Figure~\ref{fig:macs0647jd_slit_spectra} shows 2D and 1D spectra of JD1 (top right) and JD2 (bottom right) along with the slitlet positions overlaid on $1.6'' \times 1.6''$ NIRCam color images (left). We detect several emission lines, including \OIIdw\ doublet, \NeIIIw, \HeIw, \Hdelta, \Hgamma, and \OIIIwa. However, we do not detect the spectral continuum. The positions of emission lines are indicated with orange dashed lines and blue labels. The gray shaded area represents the flux uncertainties produced by the pipeline. These uncertainties might be underestimated  \citep[see e.g.,][]{2022Ferruit, Christensen2023}. Therefore, we check it by calculating the standard deviations of the spectral fluxes after removing the emission lines ($\pm 50$~\AA). The green lines in Figure \ref{fig:macs0647jd_slit_spectra} show the standard deviation level, which is higher by a factor of $\sim 2$ from the estimated flux uncertainties from the pipeline. The red lines represent the flux uncertainties multiplied by a factor of two, which we use for our analysis in this paper. The reprocessed spectra will be publicly available on our Cosmic Spring website\footnote{\url{https://cosmic-spring.github.io}}.

\section{Analysis} \label{sec:analysis}   

Most of the emission lines that were previously detected with the NIRSpec prism data \citep[see][]{Hsiao2023b} are also detected with the current NIRSpec high-resolution data, except \CIIIwa\ which is beyond the wavelength coverage of the current data. With the high-resolution grating, we can resolve \OIIdw\ doublet for the first time at $z>8$. At the central wavelength of \OIIdw, the G395H/F290LP grating has $R\sim 2855$ \citep{Jakobsen2022} which corresponds to a rest-frame dispersion of $\sim 1.3$ \AA\ per spectral resolution element. Besides this, \HeIw\ is separated from \NeIIIw, which were blended in the NIRSpec prism spectra. We describe the emission line analysis in this section.  

\subsection{Stacking and Emission Line Measurements} 
\label{sec:line_fitting}
 
We stack (i.e.,~sum) the spectra of JD1 and JD2 to get a good overall S/N ratio. The stacked spectrum has a total lensing magnification ($\mu$) of $13.3$ ($8.0+5.3$; see \citealt{Hsiao2023b,Hsiao2023a}). We do not de-lense the spectra of JD1 and JD2 before the stacking because the true line ratios of \JD\ are expected to be preserved. We measure the emission line flux and width in the individual spectra (JD1 and JD2) and the stacked spectrum. We fit the emission lines individually with a Gaussian function using \piXedfit \footnote{The line fitting function will be included in the future release of \piXedfit} \citep{Abdurrouf2021, Abdurrouf2022, Abdurrouf2023}. The code uses the Markov Chain Monte Carlo (MCMC) method through the \emcee\ package \citep{2013Foreman-Mackey} for sampling the posterior probability distributions of free parameters in the fitting. 

First, we determine the systemic redshift by fitting the strongest line \NeIIIw\ as it also has the highest S/N ratio. The spectroscopic redshift of \JD\ has been previously determined by \citet{Hsiao2023b} using NIRSpec prism data to be $10.17$. Here, we put a stronger constraint on the redshift using NIRSpec high-resolution data. We obtain $z=10.1672_{-0.0003}^{+0.0003}$, $10.1685_{-0.0004}^{+0.0004}$, and $10.1674^{+0.0002}_{-0.0002}$ for JD1, JD2, and stacked spectrum, respectively. For each spectrum, we use the systemic redshift to estimate the observed central wavelength of the other emission lines and set the initial positions of the MCMC walkers. To better constrain the FWHM of the lines, we use the FWHM posterior distribution function derived from the fitting of \NeIIIw\ as an additional prior. For \OIIdw\ doublet, we set the two lines to have the same FWHM, while the value is allowed to vary during the fitting process.  

For each emission line, we fit the spectrum region within $\pm 100$~\AA\ around the expected peak wavelength to minimize the influence of noise. We estimate the continuum level by taking the median of fluxes within the fitting region after excluding the emission line (i.e.,~masked around $\pm 40$~\AA\ from the expected peak wavelength). We then subtract the continuum from the spectrum and fit the emission line. We scale up the flux uncertainties by a factor of two to match the overall level of the standard deviation (see Figure~\ref{fig:macs0647jd_slit_spectra}). 

We do not use the measured continuum fluxes from the NIRSpec prims spectra because of the differences in the slitlet position and coverage. In \citet{Hsiao2023b}, the measurements of emission lines were performed to a stacked spectrum, which is summed of the JD1 and JD2 spectra from two observations (Obs 21 and Obs 23). In Appendix~\ref{sec:lines_nirspec_prism}, we perform emission line measurements of the individual NIRSpec prism spectra and the stacked one using the updated reductions and fitting method.         

\subsection{Measured Emission Line Properties}
\label{sec:line_properties}

\begin{figure*}[ht]
\centering
\includegraphics[width=1.0\textwidth]{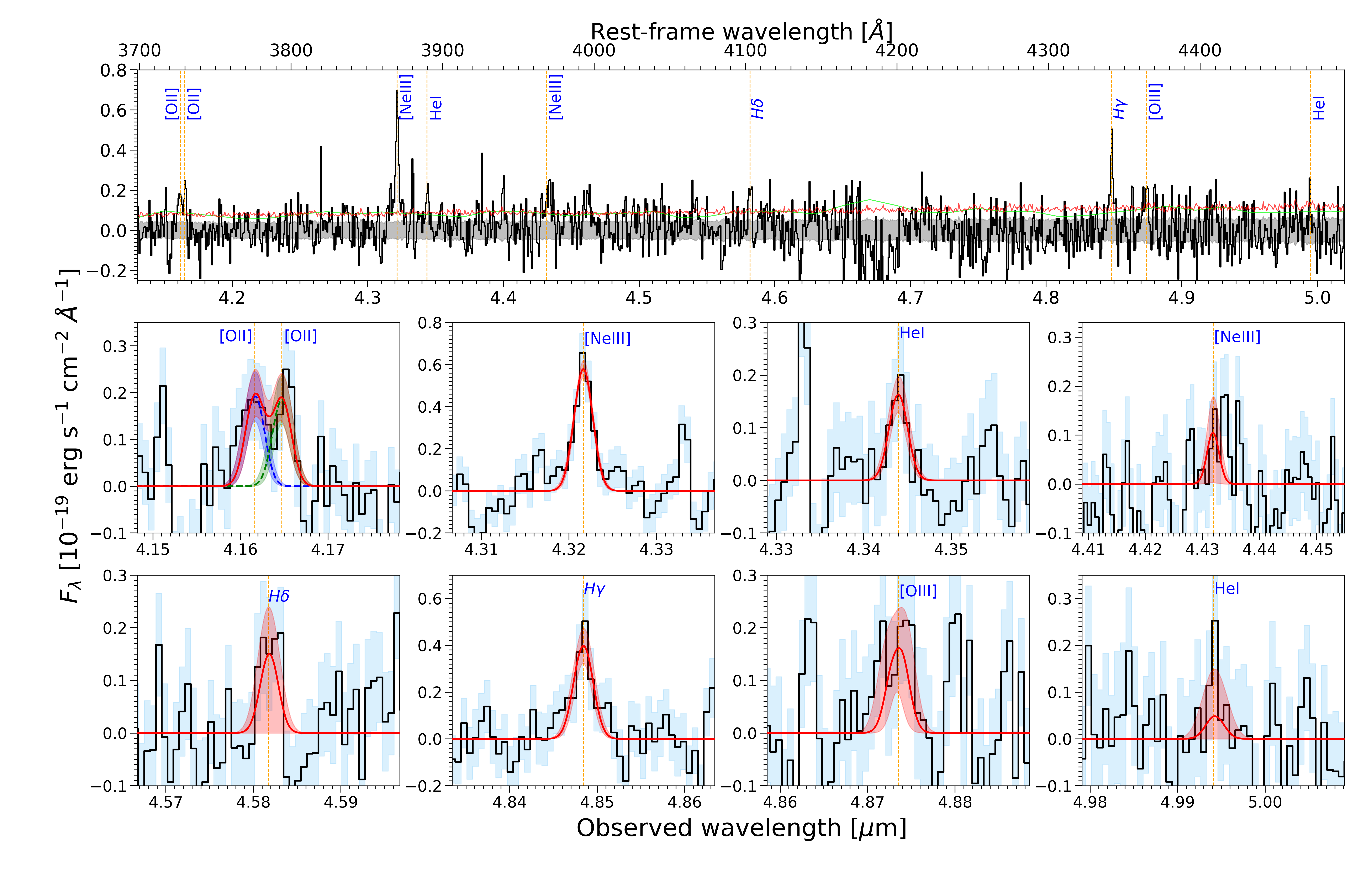}
\caption{\textit{Top}: Fitting results of the emission lines in the stacked spectrum. The gray shaded area shows the original flux uncertainties, while the red line represents a factor of two scale-up, which is used in the fitting. This matches with the overall level of standard deviations as shown by the green line. The detected emission lines are indicated with orange vertical dashed lines and blue labels. \textit{Middle and bottom}: cutouts showing the detected emission lines and the best-fit Gaussian model drawn from the posteriors obtained from the MCMC fitting. The best-fit models are shown by the red lines (median) and red-shaded areas (range between 16th and 84th percentiles). The best fits of the individual \OIIdw\ lines are shown in blue and green, while the total is shown in red. The flux uncertainties in the cutouts have been scaled up by a factor of two.}
\label{fig:spectrum_stack}
\end{figure*}

The results of the emission-line fitting for the stacked spectrum is shown in Figure~\ref{fig:spectrum_stack}, while the measured line properties are summarized in Table~\ref{tab:lines_tack}. For comparison, we also show the fitting results of the JD1 and JD2 spectra in Appendix~\ref{sec:fit_results_jd1_jd2} (see Figures~\ref{fig:spectrum_jd1} and~\ref{fig:spectrum_jd2}). 
In Figure~\ref{fig:spectrum_stack}, the first row shows the full spectrum, while cutouts in the second and third rows show fitting results of the emission lines. The red line and shaded area in each cutout show the Gaussian model drawn from the median posteriors and the range between the 16th and 84th percentiles, respectively. For the \OIIdw\ doublet, the best fits of the individual lines are shown by the blue and green dashed lines, while the red line shows the total. The flux uncertainties shown in the cutout spectra have been scaled up by a factor of two.

We detect and resolve \OIIdw\ doublet with S/N of $\sim 3-4$ on the individual lines. The flux ratio between the two lines, \ROII$\equiv$\OIIratio, is $0.91^{+0.29}_{-0.27}$ for the stacked spectrum. For comparison, the line ratios of JD1 and JD2 are $0.71^{+0.32}_{-0.22}$ and $1.04^{+0.27}_{-0.30}$, respectively. 
The flux ratios are consistent within their uncertainties, though the median values are slightly different. This is likely caused by the noise and the different slit coverage on JD1 and JD2 (see Figure~\ref{fig:macs0647jd_slit_spectra}). The stacked \OIIdw\ has an observed FWHM of $183_{-15}^{+15}$ \kms, or $150_{-19}^{+18}$ \kms\ intrinsic after correcting for instrumental line broadening of $105$ \kms.

The strongest line \NeIIIw\ is detected with a high S/N of $11.2$ and $6$ in JD1 and JD2, while it is $16.6$ in the stacked spectrum. The observed systemic FWHM from the stacked \NeIIIw\ is $176_{-14}^{+16}$ \kms, or $144_{-17}^{+19}$ \kms\ intrinsic after correcting for instrumental broadening of $101$ \kms. The good detection of \NeIIIw\ and \OIIdw\ allows us to measure Ne3O2$\equiv$\NeIIIw/\OIIdw\ ratio. From the NIRSpec high-resolution data, we obtain Ne3O2 of $2.1_{-0.4}^{+0.7}$, $0.7_{-0.2}^{+0.2}$, and $1.2_{-0.1}^{+0.2}$ for JD1, JD2, and stacked spectrum, respectively. We also determine Ne3O2 from NIRSpec prism data (see Appendix~\ref{sec:lines_nirspec_prism}) and obtain $1.2_{-0.4}^{+0.6}$, $1.0_{-0.2}^{+0.3}$, $1.1_{-0.1}^{+0.2}$, $1.7_{-0.3}^{+0.4}$ for the JD1 Obs 21, JD1 Obs23, JD2 Obs21, JD2 Obs23, respectively. Again, the discrepancies can be caused by the difference in the slit coverage and noise (in the case of the high-resolution data). Recent studies of high redshift ($3\lesssim z \lesssim 9$) galaxies using JWST/NIRSpec data measured Ne3O2 in the range of $\sim 0.2-3$ \citep[e.g.,][]{Tang2023, Reddy2023, Cameron2023, Sanders2024, Backhaus2024}, in agreement with the values determined for \JD.    

\begin{deluxetable*}{lccccc}
\tablecaption{\label{tab:lines_tack}Measured Emission Line Properties of the Stacked Spectrum
}
\tablewidth{\columnwidth}
\tablehead{
\colhead{Emission Line} &
\colhead{Rest wavelength} &
\colhead{Observed wavelength} &
\colhead{Observed Flux$^{a}$} &
\colhead{S/N} &
\colhead{Observed FWHM$^{b}$}
\vspace{-0.07in}\\
\colhead{} &
\colhead{(\AA)} &
\colhead{($\mu$m)} &
\colhead{($10^{-20}$ \cgsfluxunits)} &
\colhead{} &
\colhead{(\kms)}
}
\startdata
\OII
& 3727.10
& $4.16165_{-0.00039}^{+0.00035}$
& $55_{-14}^{+14}$ 
& $3.9$ 
& $183_{-15}^{+15}$\\
\OII
& 3729.86
& $4.16473_{-0.00039}^{+0.00036}$
& $53_{-14}^{+14}$ 
& $3.9$ 
& $183_{-15}^{+15}$\\
\NeIII
& 3869.86
& $4.32164_{-0.00007}^{+0.00007}$ 
& $157_{-9}^{+10}$
& $16.6$
& $176_{-14}^{+16}$\\
\HeI
& 3889.75 
& $4.34396_{-0.00030}^{+0.00028}$
& $45_{-9}^{+9}$ 
& $5.0$ 
& $173_{-15}^{+15}$\\
\NeIII
& 3968.59 
& $4.43192_{-0.00050}^{+0.00025}$
& $29_{-29}^{+21}$ 
& $1.4$ 
& $177_{-17}^{+37}$\\
\Hdelta
& 4102.89 
& $4.58172_{-0.00126}^{+0.00061}$ 
& $45_{-35}^{+23}$
& $1.8$ 
& $162_{-11}^{+15}$ \\
\Hgamma
& 4341.69
& $4.84842_{-0.00027}^{+0.00025}$
& $113_{-20}^{+20}$
& $5.7$
& $162_{-13}^{+13}$\\
\OIII
& 4364.44 
& $4.87352_{-0.00113}^{+0.00085}$
& $58_{-19}^{+18}$ 
& $3.1$ 
& $159_{-14}^{+13}$\\
\HeI
& 4472.74 
& $4.99409_{-0.00560}^{+0.00151}$
& $24_{-18}^{+23}$ 
& $1.3$ 
& $151_{-15}^{+13}$
\enddata
\tablenotetext{a}{Measured fluxes from the observed spectrum that have not been corrected for a lensing magnification. The stacked spectrum has the total magnification of $\mu=13.3$ ($8.0+5.3$).}
\tablenotetext{b}{Observed FWHM before correcting for the instrumental broadening.}
\end{deluxetable*}

\section{Result and Discussion} \label{sec:result_discuss}

\subsection{Measurement of Electron Density} 
\label{sec:measure_ne}

\begin{figure}[ht]
\centering
\includegraphics[width=0.48\textwidth]{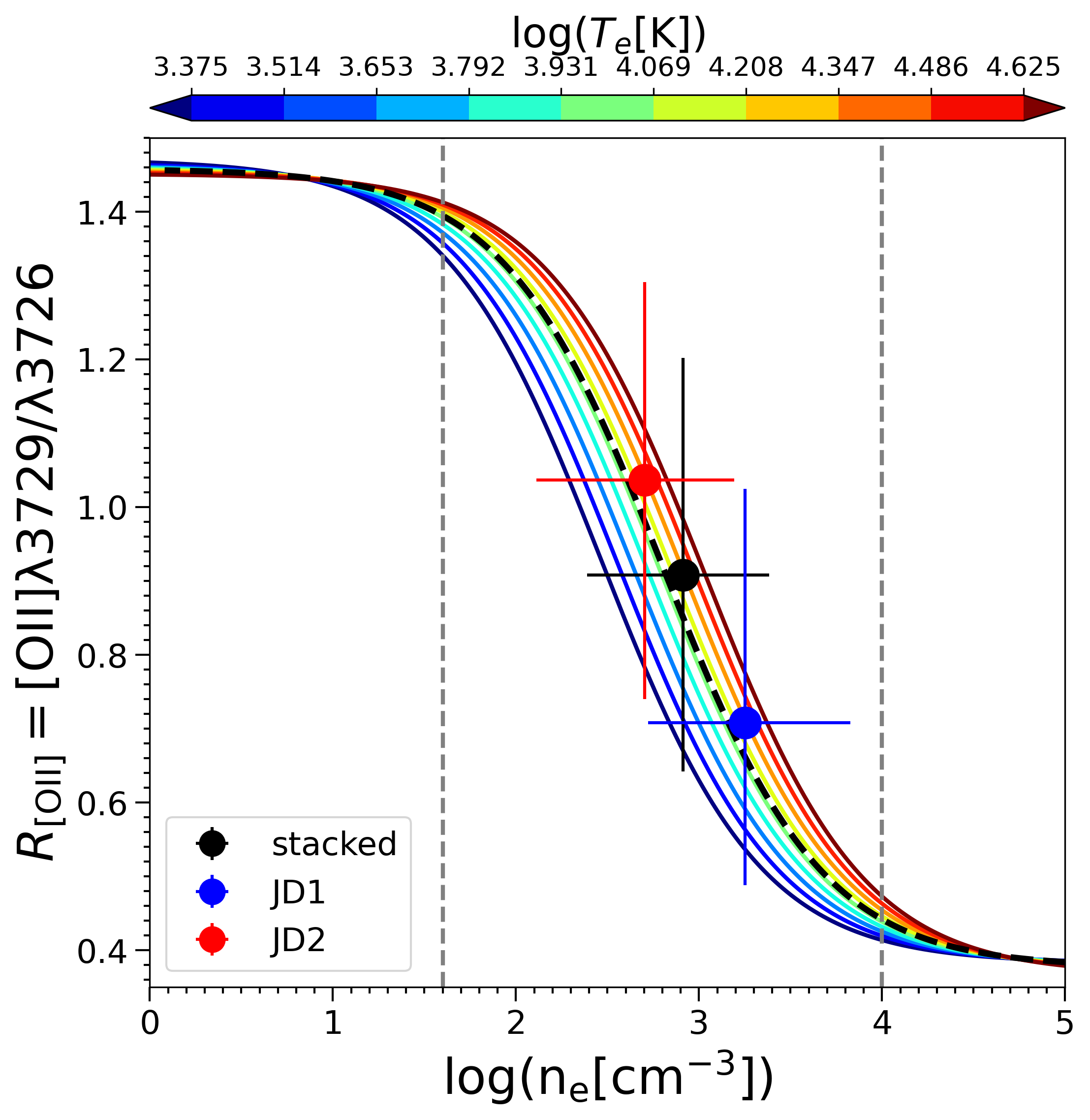}
\caption{The relationship between \OIIratio\ and the electron density drawn from the photoionization models generated by \citet{Kewley2019_a}. The curves for different temperatures are shown with different colors. Equations~\ref{eq:ne_form} and~\ref{eq:poly_function} give the formulas to reconstruct these curves. The black dashed line represents the model from \citet{Sanders2016}, which perfectly matches the \citet{Kewley2019_a} model for $T_{e}=10000$ K. The vertical gray dashed lines mark the range of electron densities for which the line ratio is a useful density diagnostic according to \citet{Kewley2019_a}. Given the \OIIratio\ line ratio and measured $T_{e}=17000$ K \citep{Hsiao2024}, the estimated $n_{e}$ of JD1, JD2, and the stacked spectrum are shown by the red, blue, and black circles, respectively.}
\label{fig:ne_vs_R}
\end{figure}

We measure \edense\ from the line ratio \ROII\ based on the photoionization models adopted from \citet{Kewley2019_a}, which were generated using the MAPPINGS version 5.1 photoionization code \citep{Binette1985, Sutherland1993, Dopita2015}. We perform interpolations to the model grids from \citet{Kewley2019_a} and construct a functional formula for converting \ROII\ to \edense\ that takes into account the electron temperature ($T_{e}$). We describe this in detail in Appendix~\ref{sec:model_interpolation}.     

Figure~\ref{fig:ne_vs_R} shows the model $R_{\rm [OII]}(n_{e})$ curves for 9 values of $T_{e}$ calculated using the constructed formula (Equations~\ref{eq:ne_form} and~\ref{eq:poly_function}). Different electron temperatures are indicated with different colors. The vertical gray dashed lines mark the range of electron densities for which the line ratio is a useful density diagnostic according to \citet{Kewley2019_a}. The black dashed line shows the relation from \citealt{Sanders2016} (Eq.~6 therein). It perfectly matches the curve from \citet{Kewley2019_a} for $T_{e}=10000$ K, which was indeed assumed in \citet{Sanders2016}. 

\begin{figure*}[ht]
\centering
\includegraphics[width=1.0\textwidth]{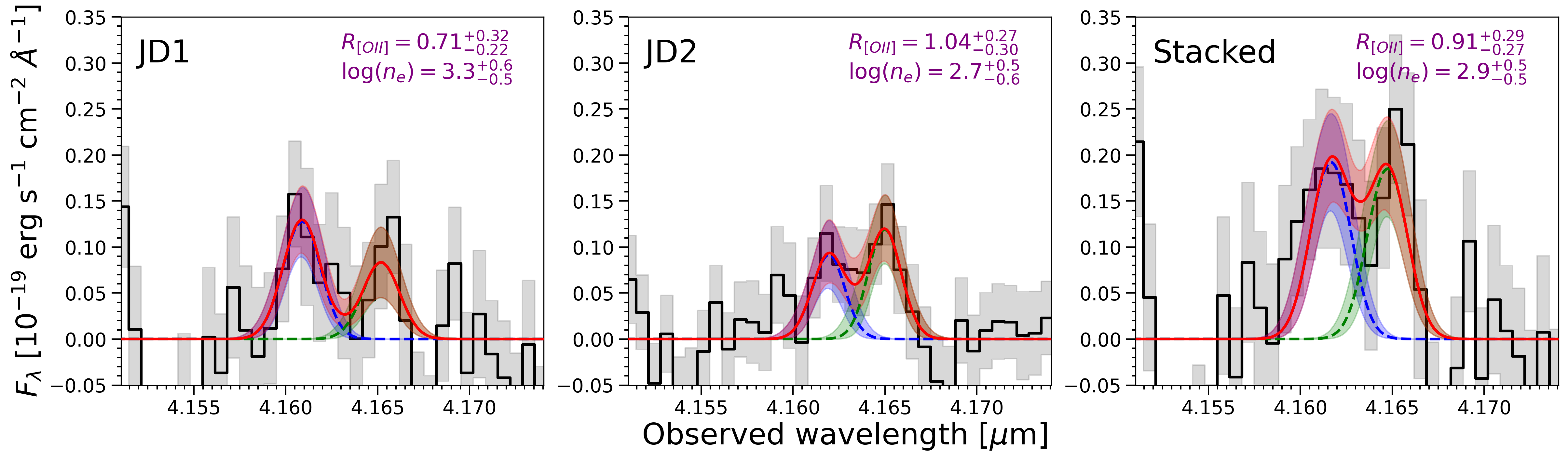}
\caption{Resolved \OIIdw\ doublet of JD1 (left), JD2 (middle), and the stacked spectrum (right). The two lines are resolved in the three spectra. Green dashed and blue dashed lines are the best-fit Gaussian of \OIIwa\ and \OIIwb\, respectively, while the solid red line represents the total. Derived \ROII\ of JD1, JD2, and stacked spectrum are $0.71^{+0.32}_{-0.22}$, $1.04^{+0.27}_{-0.30}$, and $0.91^{+0.29}_{-0.27}$, respectively. Assuming $T_{e}=17000$ K \citep{Hsiao2024}, we estimate \edense\ of $\log(n_{e})=3.3^{+0.6}_{-0.5}$, $2.7^{+0.5}_{-0.6}$, and $2.9^{+0.5}_{-0.5}$, respectively.}
\label{fig:OII_doublet}
\end{figure*}

We use the new formula for calculating the \edense\ of \JD\ system. From our fitting (see Sections~\ref{sec:line_fitting} and~\ref{sec:line_properties}), we obtain \ROII\ of $0.71^{+0.32}_{-0.22}$, $1.04^{+0.27}_{-0.30}$, and $0.91^{+0.29}_{-0.27}$ for the JD1, JD2, and stacked spectrum, respectively. Assuming $T_{e}=17000$ K for \JD\ as calculated in a companion paper \citep{Hsiao2024}, we obtain electron densities of $\log(n_{e}/\rm{cm}^{-3})=3.3^{+0.6}_{-0.5}$, $2.7^{+0.5}_{-0.6}$, and $2.9^{+0.5}_{-0.5}$, respectively. In Figure~\ref{fig:ne_vs_R}, those values are indicated with blue, red, and black circles, respectively. With the current measured \ROII, we check that changing the assumptions of electron temperature from $T_{e}=10000$ to $30000$ K increases \edense\ by a factor of $\sim 1.7$. 

We also estimate \edense\ using \pyneb\ \citep{Luridiana2015} task \texttt{getCrossTemDen} and combine measured \ROII\ from the stacked spectrum and \OIII\,$\lambda$5007/$\lambda$4363~=~$40 \pm 5$ line ratio measured in a companion paper \citep{Hsiao2024}. We obtain $\log(n_{e})=2.9_{-0.4}^{+0.3}$, which is consistent with the value derived above.  

We highlight the fitting results of \OIIdw\ in Figure~\ref{fig:OII_doublet}. Even though JD1 and JD2 are lensed images of the same galaxy, there is an opposite trend of \ROII\ between JD1 and JD2 if we examine the median values. However, they are consistent within the range given their uncertainties. With the MCMC method performed here, we propagate the uncertainty from the spectra to the posterior probability distribution of \edense\ in a self-consistent way. We show the posterior distributions of \edense\ (along with the fluxes of the two lines in \OIIdw\ doublet, and FWHM) in Appendix~\ref{sec:corner_plot} (Figure~\ref{fig:corner_plots}). The posteriors of these parameters are converged, indicating a good fitting. The apparent cut (i.e.,~envelope) in the 2D posterior between the two line fluxes is due to the range given by the \ROII\ within which \edense\ is physically realistic. From the Equations~\ref{eq:ne_form} and~\ref{eq:poly_function}, this range is between $0.3729$ and $1.4556$ for $T_{e}=17000$ K. We note that this range is not imposed from the beginning (i.e.,~in the fitting process). 

\ROII\ is also a good indicator of the ISM pressure, which is defined as the pressure within the ionized gaseous nebulae surrounding young hot
stars \citep[e.g.,][]{Kewley2019_a}. Using the photoionization models of \citet[][Figure~10 therein]{Kewley2019_a} and assuming $12+\log(\text{O}/\text{H})=7.79\pm0.09$ \citep{Hsiao2024}, we estimate ISM pressure of $\log(P/\text{k})=7.3_{-0.3}^{+0.4}$.

\subsection{Redshift Evolution of Electron Density}
\label{sec:evolution_ne}

\begin{figure*}[ht]
\centering
\includegraphics[width=0.75\textwidth]{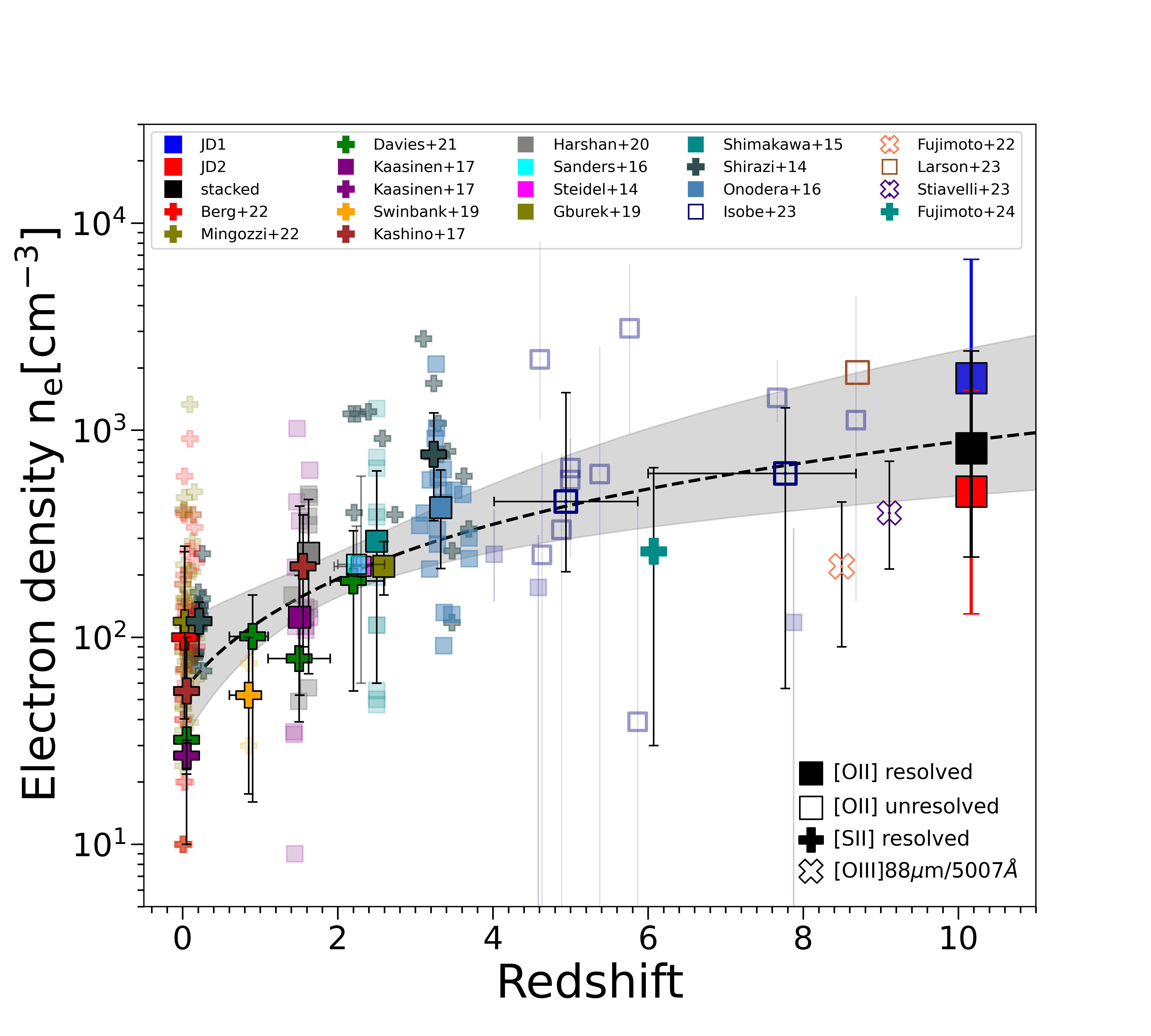}
\caption{Electron densities of galaxies measured from $z = 0$ to 10. Our measurements for JD1, JD2, and the stacked spectrum are shown by the blue, red, and black squares, respectively. The rest of the \edense\ measurements are compiled from the literature. For local galaxies at $z\lesssim 0.2$: \citet{Berg2022}, \citet{Davies2021}, \citet{Kashino2017}, \citet{Kaasinen2017}, \citet{Shirazi2014}, and \citet{Mingozzi2022}. For $0.2\lesssim z\lesssim 4$: \citet{Davies2021}, \citet{Swinbank2019}, \citet{Kashino2017}, \citet{Kaasinen2017}, \citet{Harshan2020}, \citet{Sanders2016}, \citet{Steidel2014}, \citet{Shimakawa2015}, \citet{Gburek2019}, \citet{Onodera2016}, and \citet{Shirazi2014}. For $4\lesssim z\lesssim 9$: \citet{Isobe2023}, \citet{Larson2023}, \citet{Fujimoto2022}, \citet{Stiavelli2023}, and \citet{Fujimoto2024}. Different colors represent different studies, while different symbols represent various emission-line diagnostics used to derive \edense. Overall, there are three employed emission-line diagnostics to derive \edense: \ROII\ (square), \RSII\ (plus), and an indirect method using \OIII\,$\lambda$88$\mu$m/$\lambda$5007 ratio (cross). While \OIIdw\ doublet is resolved in the majority of the studies (indicated with closed squares), it is not resolved in some studies (indicated with open squares) because of the insufficient spectral resolution. For studies that reported \edense\ of the individual galaxies, we show those with transparent symbols, in addition to the median values shown with thick symbols. The black dashed line represents the best-fit power-law function in the form of $n_{e} = A\times (1+z)^{p}$ with $A=54^{+31}_{-22}$ cm$^{-3}$ and 
$p=1.2 \pm 0.4$. The gray shaded area shows 2$\sigma$ uncertainty around the best-fit line.}
\label{fig:electron_density_vs_z}
\end{figure*}

Here, we put the measured electron density of \JD\ into the context of galaxy evolution by comparing it with the measurements at lower redshifts. We compile measured electron densities at $z=0-9$ reported in the literature and plot them in Figure~\ref{fig:electron_density_vs_z}. Different symbols represent various emission-line diagnostics used in deriving \edense, while different colors represent different studies. Three emission-line diagnostics have been used in the literature to derive \edense: \ROII\ (square), \RSII\ (plus), and an indirect method using \OIII\,$\lambda$88$\mu$m/$\lambda$5007 ratio (cross). Some observations do not resolve \OIIdw\ doublet because of the insufficient spectral resolution. We indicate the results from those studies with open squares. For studies that reported \edense\ of the individual galaxies, we show those values with transparent symbols, in addition to the median values shown with thick symbols. The measured \edense\ of JD1, JD2, and stacked spectrum are shown with the blue, red, and black squares, respectively. The rest of the data points are listed in the following, classified based on the redshift range.  

\begin{itemize}
\item \underline{$z\lesssim 0.2$}:
\citet{Berg2022} measured \edense\ of 45 local star-forming galaxies at $0.002<z<0.182$ as part of the CLASSY survey \citep{Berg2022,James2022} which aims to study the ISM properties of local high-$z$ analogs. They derived \edense\ based on \RSII\ $=$\SIIratio\ and obtained a wide range of \edense\ $\sim 10 - 900$ cm$^{-3}$, with a median of $100$ cm$^{-3}$. 
Using the same CLASSY sample, \citet{Mingozzi2022} explored multiple UV and optical diagnostics for measuring the properties of ISM, including \edense, in which they applied multiple line ratios representing low- and high-ionization regions. We take \edense\ measured based on \RSII\ whose $T_{e}$ is derived based on \OIII\,$\lambda$4363/$\lambda$5007.
\citet{Davies2021} measured \edense\ of galaxies in a wide redshift range by stacking the spectra of the galaxies in four redshift bins: $z\lesssim 0.1$ (471), $0.6\lesssim z\lesssim 1.1$ (39), $1.1\lesssim z\lesssim 1.9$ (36), and $1.9\lesssim z\lesssim 2.6$ (65). The data for local galaxies are taken from SAMI survey \citep{Croom2012}. They estimated $n_{e}=32_{-9}^{+4}$ cm$^{-3}$ from \RSII. 
\citet{Kashino2017} and \citet{Kaasinen2017} measured \edense\ of galaxies at $z\sim 1.5$ and $z\sim 0$. For the local sample, they used data from the SDSS Data Release 7 \citep{Abazajian2009}, specifically the MPA-JHU catalogs \citep{Kauffmann2003,Brinchmann2004}. For the local sample, \citet{Kaasinen2017} obtained a median of \edense~$26.8_{-0.2}^{+0.2}$ cm$^{-3}$, while \citet{Kashino2017} obtained \edense\ in the range of $\sim 10-100$ cm$^{-3}$.   
\citet{Shirazi2014} studied the ISM properties in 14 star-forming galaxies at $2.6<z<3.4$ and local galaxies matched in stellar mass and sSFR. They used emission lines from the MPA-JHU catalogs for the local sample, and estimated \edense\ based on \RSII, finding a range of $\sim 60 - 250$ cm$^{-3}$.      

\item \underline{$0.2\lesssim z \lesssim 4$}: 
\citet{Davies2021} used data from the KMOS$^{\rm 3D}$ survey \citep{Wisnioski2015} to measure \edense\ of galaxies in 3 high-$z$ bins (mentioned above). Based on the \RSII\ of the stacked spectra, they estimated $n_{e}=101^{+59}_{-85}$, $79^{+120}_{40}$, and $187^{+140}_{-132}$ cm$^{-3}$ for the samples at $z \sim 0.9$, $\sim 1.5$, and $\sim 2.2$, respectively. 
\citet{Swinbank2019} used data from the KROSS survey \citep{Stott2016} and performed stacking on the spectra of 529 star-forming galaxies at $0.6<z<1.0$. They estimated \edense\ from \RSII\ in the stacked spectrum and decomposed it into broad-line and narrow-line components. They estimated \edense\ $=30_{-20}^{+40}$ and $75_{-50}^{+55}$ cm$^{-3}$ for the narrow-line and broad-line components, respectively. We take the mean of the two for our analysis. 
\citet{Kashino2017} used data from the FMOS-COSMOS survey \citep{Silverman2015} to study the ISM properties of galaxies at $1.4\lesssim z\lesssim 1.7$. They measured an average \edense\ $=220_{-130}^{+170}$ cm$^{-3}$ based on \RSII.
\citet{Kaasinen2017} measured \edense\ of 21 star-forming galaxies at $1.4\lesssim z\lesssim 1.7$ using data from the COSMOS-[OII] survey. The estimated \edense\ based on \ROII\ ranges from $\sim 9$ to $1000$ cm$^{-3}$, with a median of $114^{+28}_{-27}$ cm$^{-3}$.
\citet{Harshan2020} used data from the ZFIRE survey \citep{Nanayakkara2016} to calculate \edense\ of $z\sim 1.5$ galaxies, of which 6 are in the cluster environment and 2 are in the field. 
They found \edense\ in the range of $\sim 57 - 490$ cm$^{-3}$ for the cluster galaxies and $\sim 49 - 160$ cm$^{-3}$ for the field galaxies.
\citet{Sanders2016} measured \edense\ of 97 galaxies at $2.0<z<2.6$ using data from the MOSDEF survey \citep{Kriek2015}. They obtained a median \edense\ $=225^{+119}_{-4}$ cm$^{-3}$ based on \ROII.
\citet{Steidel2014} estimated \edense\ $=220^{+380}_{-160}$ cm$^{-3}$ from \ROII\ of a stacked spectrum of 113 galaxies at $2.0<z<2.6$. 
\citet{Shimakawa2015} measured \edense\ of 14 \Halpha-emitting galaxies at $z\sim 2.5$ based on \ROII. They obtained a median of \edense\ $\sim 290$ cm$^{-3}$. 
\citet{Gburek2019} measured \edense\ $\sim 220$ cm$^{-3}$ in a lensed dwarf galaxy at $z=2.59$ based on \ROII. 
\citet{Onodera2016} derived \edense\ of star-forming galaxies at $3\lesssim z\lesssim 3.7$ based on \ROII, finding a median $\sim 420$ cm$^{-3}$. 
\citet{Shirazi2014} obtained \edense\ $\sim 110 - 2700$ cm$^{-3}$ in star-forming galaxies at $2.6<z<3.4$ based on \RSII.  

\item \underline{$4\lesssim z\lesssim 9$}: 
\citet{Isobe2023} estimated \edense\ of 14 star-forming galaxies at $4.02<z<8.68$ using data from public JWST surveys that used the NIRSpec medium- and high-resolution spectroscopy. They observed \OIIdw\ doublet resolved in only three galaxies, but blended in the other galaxies. They carefully modeled the line spread function (LSF) of the NIRSpec to deconvolve the \OIIdw\ doublets. They obtained \edense\ range of $\sim 39-3100$ cm$^{-3}$. \citet{Larson2023} studied the ISM properties in a galaxy CEERS\_1019 at $z=8.679$ using NIRSpec medium-resolution data. They carefully fit an unresolved \OIIdw\ doublet with two Gaussian models and derived \edense\ $1.9 \pm 0.2 \times 10^{3}$ cm$^{-3}$.    
\citet{Fujimoto2022} measured \edense\ of a galaxy at $z=8.496$ using indirect method based on the \OIII\,88$\mu$m/$\lambda$5007 line ratio and obtained \edense\ $=220_{-130}^{+230}$ cm$^{-3}$. 
Using the same method as \citet{Fujimoto2022}, \citet{Stiavelli2023} estimated \edense\ of MACS1149-JD1 at $z=9.11$ to be $\log(n_{e})=2.60_{-0.27}^{+0.25}$.  
With the \SIIdw\ doublets, \citet{Fujimoto2024} evaluated \edense\ of a sub-$L^{\star}$ main-sequence lensed galaxy at $z=6.07$, {\it the Cosmic Grapes}, to be $n_{e}=260^{+400}_{-230}$~cm$^{-3}$.
\end{itemize}  

As shown in Figure~\ref{fig:electron_density_vs_z}, there is an indication of an increasing trend of $n_{e}$ with redshift from $z=0$ to $10$, especially for the median values in the redshift bins (thick symbols). To get a functional form out of the evolutionary trend, we fit the data with a power-law function in the form of $n_{e} = A\times (1+z)^{p}$ using the MCMC method. For the literature that reports electron densities of multiple individual galaxies in a redshift bin, we only use the median value for the fitting. For \JD, we only use the electron density derived from the stacked spectrum. We obtain $A=54^{+31}_{-22}$ cm$^{-3}$ and $p=1.2^{+0.4}_{-0.4}$. The median function obtained from the fitting result is shown with the black dashed line, while the gray shaded area shows 2$\sigma$ uncertainty. The fitting result can track the bulk of the data distribution quite well. Best-fit $A$ suggests an electron density of $\sim 50$ cm$^{-3}$ in the local universe. The \edense\ of JD1, JD2, and stacked spectrum falls within the range of the best-fit line and 2$\sigma$ uncertainty. While the median \edense\ values follow the increasing trend with redshift, we note the relatively wide scatter of the individual \edense\ values around the best-fit function. Some CLASSY galaxies have \edense\ that reach $\sim 10^3$~cm$^{-3}$. This can be because those galaxies are local high-$z$ analogs.

Previous studies have also hinted at a redshift evolution of electron density in the form of a power-law function, $n_{e}\propto (1+z)^{p}$ with $p\sim 1-2$ \citep[e.g.,][]{Davies2021, Isobe2023}. \citet{Isobe2023} compiled electron densities reported from the literature at $z\lesssim 3$ and combined with their measurements at $4\lesssim z\lesssim 9$. They found that the majority of the data lie within the range given by the power-law function with $p=1$ and $2$. They further investigate the dependence of \edense\ on the $M_{*}$, SFR, and sSFR, and found that \edense\ seems to increase toward higher redshift at a given $M_{*}$, SFR, and sSFR, suggesting that \edense\ evolution is not an implication of the evolution of those global properties. This somewhat contradicts the finding of \citet{Kaasinen2017} that shows a weak evolution of \edense\ with redshift for a given SFR, suggesting that the redshift evolution of \edense\ is mainly due to the cosmic evolution of SFR. To further investigate the driving factors of \edense\ evolution, we need more data that cover a wide range of redshift and homogeneous methods in the measurements of the physical properties, which is beyond the scope of our current study. We do not investigate the dependence of \edense\ on the global properties of galaxies because of the inhomogeneity of the methods for deriving those properties used in the literature.                         

The values $n_e \approx 10^3$~cm$^{-3}$ measured in JD1 and JD2 are in good agreement with the typical density of the ISM in the vicinity of massive stars, within star-forming regions, found in cosmological simulations of $z \sim 10$ galaxies \citep[e.g., see Fig.~12 in][]{Sugimura2024}. However, understanding the dominant driver of the redshift evolution of $n_e$ is more complicated. Although the mean density of the IGM and virialized halos scale as $(1+z)^3$, the phase of the ISM probed by the \OIIdw\ emission is ionized gas around newly formed massive stars, which is several orders of magnitude denser than the mean gas density in halos. One intriguing interpretation of the redshift evolution of $n_e$ is in terms of the mean metallicity evolution of galaxies with redshift. This suggestion is based on two observations: i) Simulations of high-redshift galaxies find that the mean density of star-forming clumps roughly scales as $\overline n \propto Z^{-1}$, a result that is a direct consequence of their higher temperatures in lower-metallicity gas clouds \citep[see Fig.~3 in][]{Garcia2023}; ii) The mean metallicity of the ISM in star-forming galaxies at redshifts $z \sim 0-3$, according to simulations \citep{Dave2011} and observations \citep{Yuan2013}, decreases with increasing redshift roughly as $Z \propto (1+z)^{-1}$, hence the relationship $n_e \propto \overline n \propto (1+z)$.

Alternatively, the \edense\ evolution in the form of $n_{e}\propto (1+z)^{p}$ may be related to the morphological evolution of galaxies which influences the density of the ISM. The increasing ISM density toward high redshift can be caused by the size evolution of galaxies, which are more compact at high redshift. The galaxy size has been shown to decrease with redshift in the form of $r \propto (1+z)^{\sim -1}$ \citep[e.g.,][]{Shibuya2015,Ono2023,Ormerod2024}. The virial radius of the dark matter halo also evolves at a similar rate \citep[e.g.,][]{Mo2002}. Assuming disk morphology, which is compatible with star-forming galaxies, the densities of stellar mass and dark matter halo mass are expected to evolve as $r^{-2} \propto (1+z)^{\sim 2}$. If \edense\ is expected to evolve following the stellar mass and halo mass densities, it evolves as $n_{e}\propto (1+z)^{\sim 2}$. Our result suggests a slightly lower power index of $1.2^{+0.5}_{-0.4}$.

\section{Summary} \label{sec:summary}

We present JWST/NIRSpec high-resolution spectroscopy (G395H/F290LP grating) of \JD, a triply-lensed galaxy system at $z=10.167$. We obtained spectra for the MACS0647--JD1 and --JD2. The new spectra detect and resolve emission lines in the rest-frame ultraviolet (UV) and blue optical, including resolved \OIIdw\ doublet, \NeIIIw, \HeIw, \Hdelta, \Hgamma, and \OIIIwa. The \OIIdw\ is resolved for the first time at $z>8$, revealing the two lines and providing an unprecedented opportunity to directly measure electron density (\edense) in a galaxy at early cosmic time. We stack the spectra of JD1 and JD2 to get a higher overall signal-to-noise ratio. We measure \edense\ from the \OIIratio\ and obtained $\log(n_{e}/\rm{cm}^{-3})=3.3^{+0.6}_{-0.5}$, $2.7^{+0.5}_{-0.6}$, and $2.9^{+0.5}_{-0.5}$ for JD1, JD2, and stacked spectrum, respectively. We compare our \edense\ measurements with those reported in the literature for galaxies in lower redshifts and study the evolutionary trend of \edense. We see a clear trend of increasing \edense\ with redshift. The redshift evolution follows a power-law function in the form of $n_{e} = A\times (1+z)^{p}$ with $A=54^{+31}_{-22}$ cm$^{-3}$ and $p=1.2^{+0.4}_{-0.4}$. This power-law form may be explained as the result of decreasing gas-phase metallicity of galaxies with increasing redshift, the morphological evolution of galaxies, or the combination of the two. It has been known that lower-metallicity star-forming clouds tend to be denser and the morphological evolution of galaxies happened in such a way that galaxies were more compact at higher redshift.


\section{Acknowledgments}
We thank the anonymous referee for providing valuable
comments that helped to improve this paper.
This work is based on observations made with the NASA/ESA/CSA 
\textit{James Webb Space Telescope} (\JWST). 
The JWST data presented in this article were obtained from the Mikulski Archive for Space Telescopes (MAST) at the Space Telescope Science Institute. The specific observations analyzed can be accessed via \dataset[DOI]{https://doi.org/10.17909/8yby-hd71}. The Association of Universities for Research in Astronomy (AURA), Inc. operates the MAST under NASA contract NAS 5-03127 for \JWST.

A and TYYH are funded by a grant for JWST-GO-01433 and JWST-GO-04246 provided by STScI under NASA contract NAS5-03127.
TYYH appreciates the support from the Government scholarship to study abroad (Taiwan).
We are grateful and indebted to the 20,000 people who worked to make \JWST\ an incredible discovery machine.
AA acknowledges support by the Swedish research council Vetenskapsr{\aa}det (2021-05559). PD acknowledges support from the NWO grant 016.VIDI.189.162 (``ODIN") and warmly thanks the European Commission's and University of Groningen's CO-FUND Rosalind Franklin program. 
M.K. was supported by the ANID BASAL project FB210003.
AZ and LJF acknowledge support by Grant No. 2020750 from the United States-Israel Binational Science Foundation (BSF) and Grant No. 2109066 from the United States National Science Foundation (NSF); by the Ministry of Science \& Technology, Israel; and by the Israel Science Foundation Grant No. 864/23.

%

\vspace{5mm}
\facilities{\textit{JWST}/NIRSpec}


\software{
\grizli\ \citep{Brammer2022},
astropy \citep{astropysoftware1,astropysoftware2,astropysoftware3},
matplotlib \citep{hunter2007},
NumPy \citep{harris20numpy},
SciPy \citep{scipy20},
STScI \textit{JWST} Calibration Pipeline (\url{jwst-pipeline.readthedocs.io}; \citealt{Rigby2023}, \citealt{bushouse_2023_10022973}),
IDL Astronomy Library (\url{idlastro.gsfc.nasa.gov}; \citealt{Landsman93}),
\piXedfit\ \citep{Abdurrouf2021,Abdurrouf2022},
\msaexp\ \citep{brammer_msaexp},
\pyneb\ \citep{Luridiana2015}.
}



\appendix

\section{Fitting results of the emission lines in the JD1 and JD2 spectra}
\label{sec:fit_results_jd1_jd2}

\begin{figure*}[ht]
\centering
\includegraphics[width=1.0\textwidth]{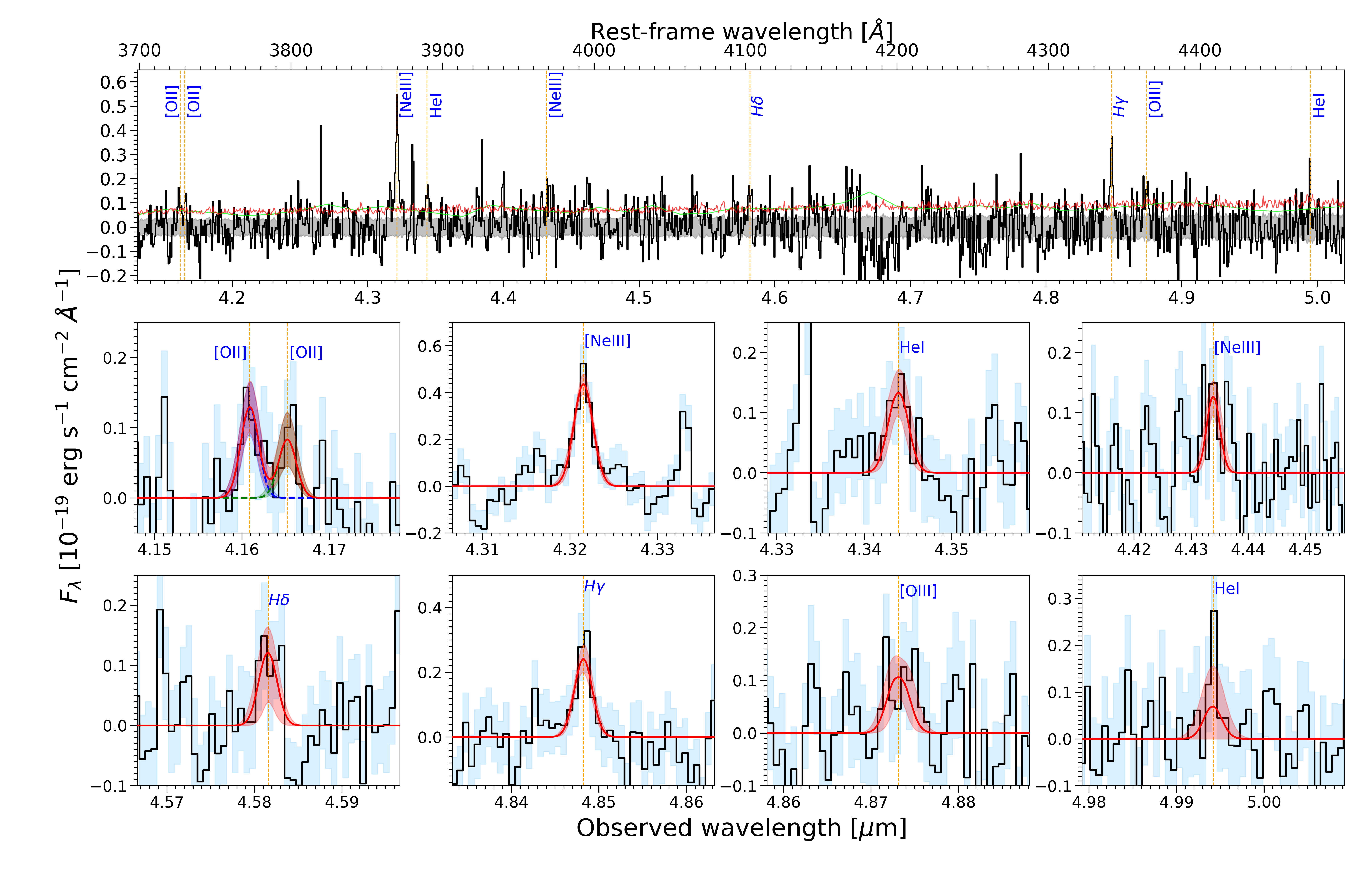}
\caption{Same as Figure~\ref{fig:spectrum_stack} but for JD1 spectrum.}
\label{fig:spectrum_jd1}
\end{figure*}

\begin{figure*}[ht]
\centering
\includegraphics[width=1.0\textwidth]{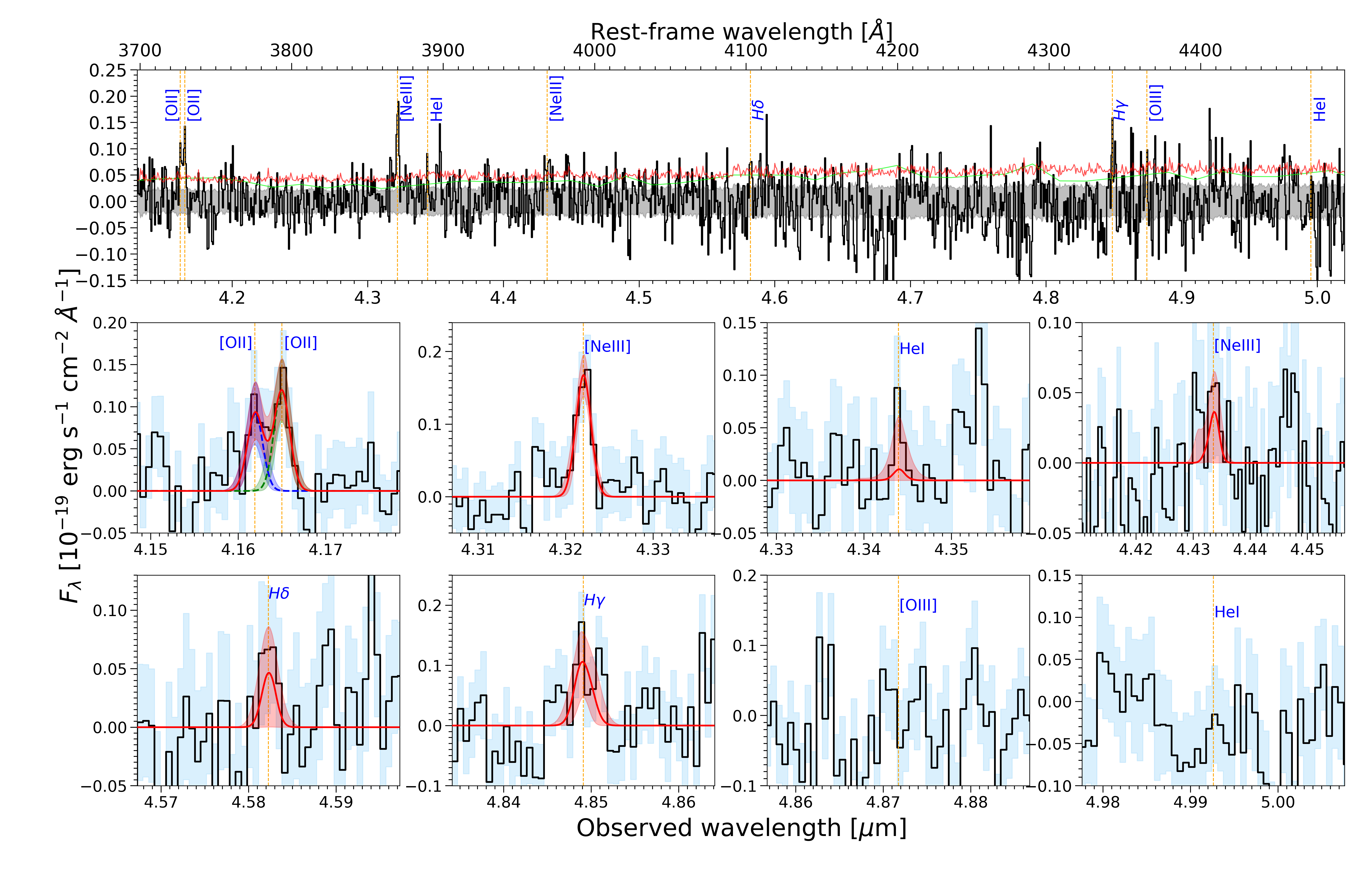}
\caption{Same as Figure~\ref{fig:spectrum_stack} but for JD2 spectrum. The fitting results of \OIIIw\ and \HeIwa\ are omitted because of the insignificant detection of the lines.}
\label{fig:spectrum_jd2}
\end{figure*}

The results of emission-line fitting for the JD1 and JD2 spectra are shown in Figures~\ref{fig:spectrum_jd1} and~\ref{fig:spectrum_jd2}, respectively. The measured properties of the emission lines are summarized in Tables~\ref{tab:lines_jd1} and~\ref{tab:lines_jd2}.

\section{General formula for the relationship between \ROII\ and \edense}
\label{sec:model_interpolation}

\begin{figure*}[ht]
\centering
\includegraphics[width=1.0\textwidth]{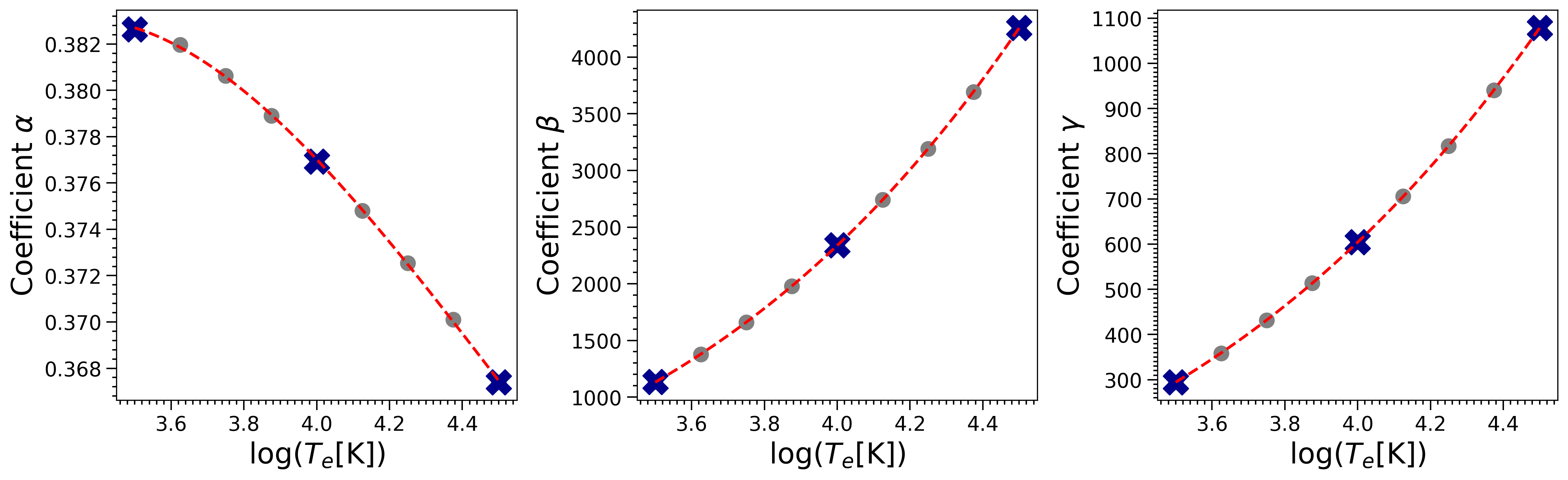}
\caption{The dependence of the coefficients in Equation~\ref{eq:ne_form} to the electron temperature ($T_{e}$; in logarithmic scale). The blue crosses represent the coefficients of the original $R_{\rm [OII]}(n_{e})$ curves from \citet{Kewley2019_a} at $\log(T_{e}/\rm{K})=3.5, 4.0,$ and $4.5$. The gray circles show the coefficients of the interpolated curves at 9 values of $T_{e}$. The red dashed lines show the best-fit cubic polynomial functions.}
\label{fig:model_coefficients_poly}
\end{figure*}

We perform interpolation to the model grids of \citet{Kewley2019_a} to construct a general functional formula for calculating \edense\ from \ROII\ given an electron temperature ($T_{e}$). The model grids of \citet{Kewley2019_a} give the relationship between \edense\ and \ROII\ for three different electron temperatures ($\log(T_{e}/\rm K)=3.5, 4.0,$ and $4.5$). The model grids were generated using the MAPPINGS version 5.1 photoionization code \citep{Binette1985, Sutherland1993, Dopita2015}. To get a functional formula out of the curves, we fit each $R_{\rm [OII]}(n_{e})$ curve with the following functional form adapted from \citet{Sanders2016}:
\begin{equation}
n_{e} = \frac{\gamma R_{\rm [OII]}-\alpha \beta}{\alpha-R_{\rm [OII]}}. 
\label{eq:ne_form}
\end{equation}

We then interpolate the $R_{\rm [OII]}(n_{e})$ curves using the cubic spline method for 9 values of $\log(T_{e})$ between $3.5$ and $4.5$. For each interpolated curve, we perform fitting with Eq.~\ref{eq:ne_form} to get the coefficients ($\alpha$, $\beta$, and $\gamma$) and investigate their dependencies with $T_{e}$. We show the relations between those coefficients and $T_{e}$ in Figure~\ref{fig:model_coefficients_poly}. The blue crosses represent the coefficients of the original curves from \citet{Kewley2019_a}, while the gray circles are from the interpolated curves. The relations can be well-fitted with a cubic polynomial function of the form
\begin{equation}
\alpha, \beta, \gamma = a + bx + cx^{2} + dx^{3}
\label{eq:poly_function}
\end{equation}
with $x=\log(T_{e})$. The best-fit cubic polynomial functions are shown as red dashed lines in Figure~\ref{fig:model_coefficients_poly} and their coefficients are summarized in Table~\ref{tab:poly_coeff}. We show models $R_{\rm [OII]}(n_{e})$ for 9 values of $T_{e}$ in Figure~\ref{fig:ne_vs_R}.

\section{Corner plots of the fitting on \OIIdw\ doublet}
\label{sec:corner_plot}

\begin{figure*}[ht]
\centering
\includegraphics[width=0.45\textwidth]{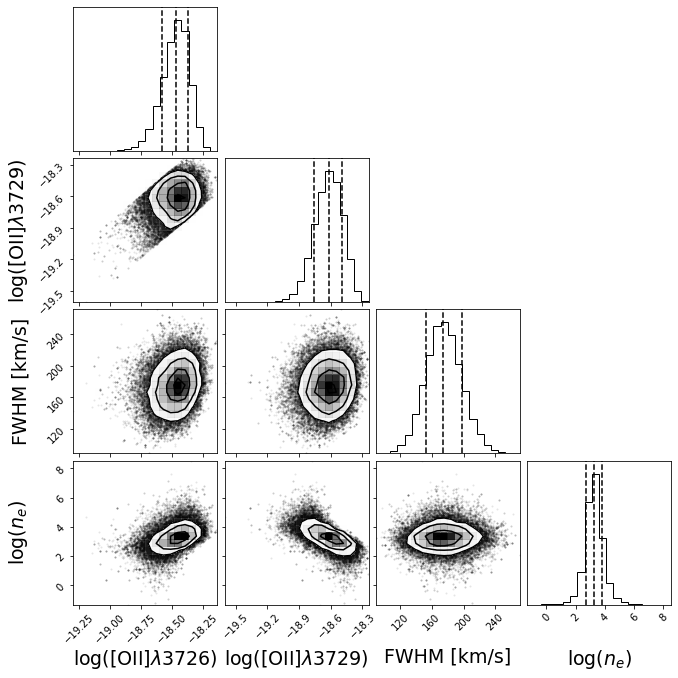}
\includegraphics[width=0.45\textwidth]{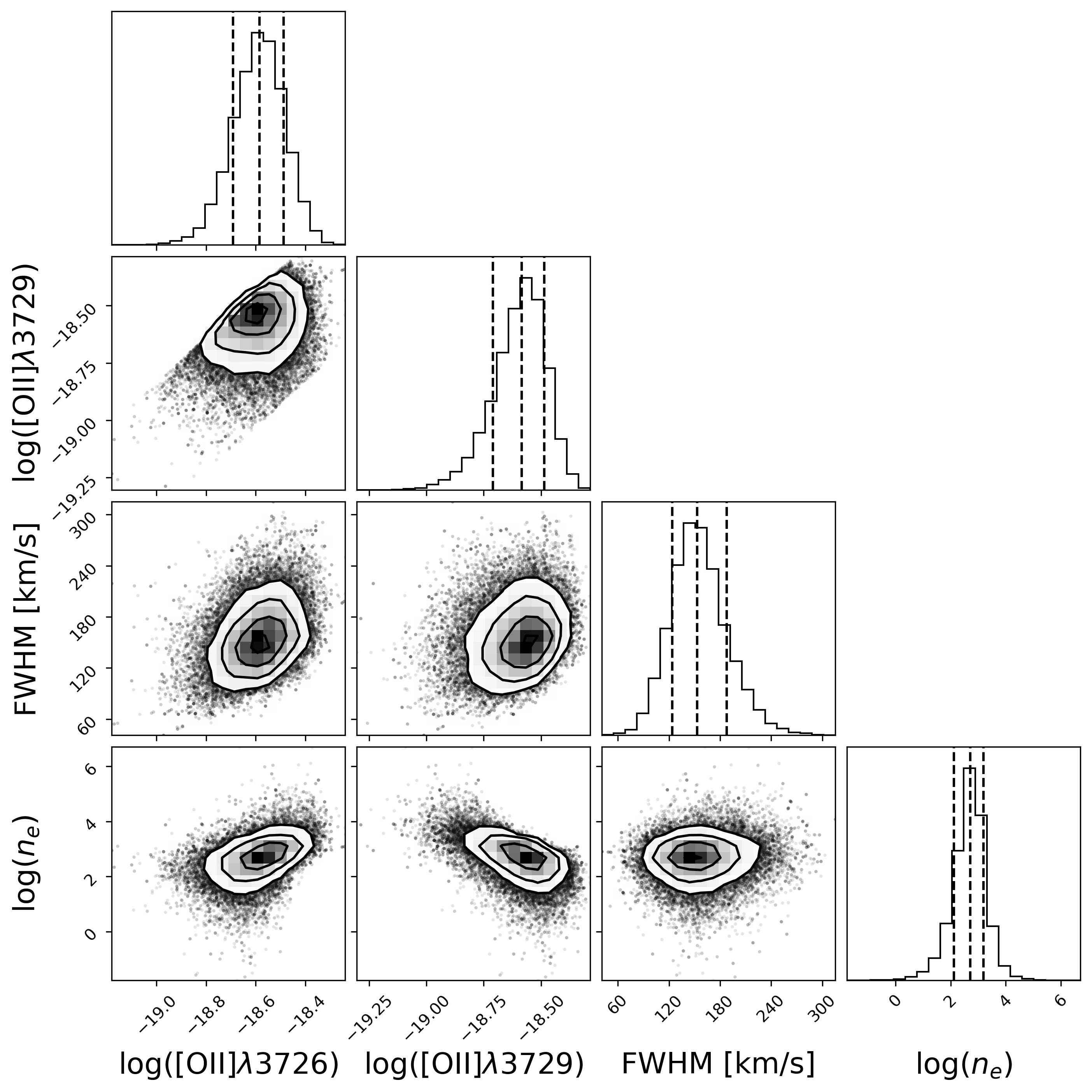}
\includegraphics[width=0.45\textwidth]{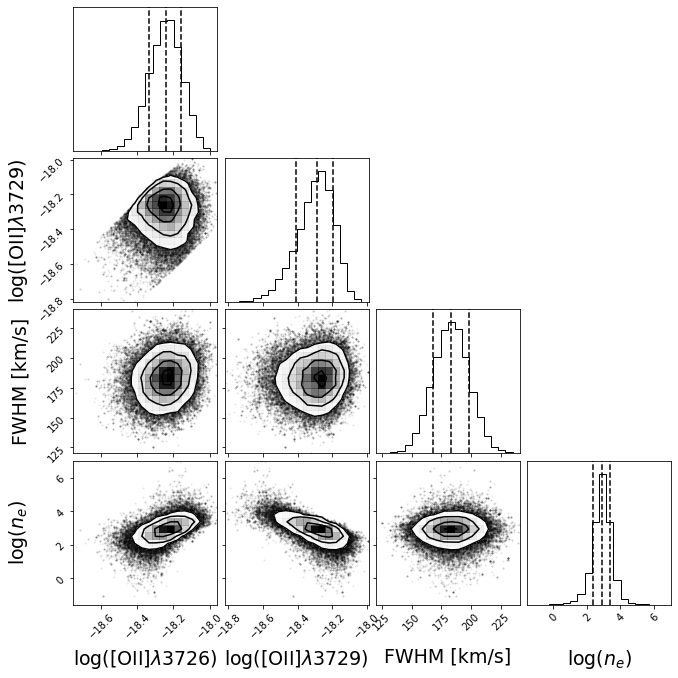}
\caption{Corner plots showing the posterior probability distributions of some parameters in the MCMC fitting of the \OIIdw\ doublet. The top left, top right, and bottom panels show the corner plots for JD1, JD2, and stacked spectrum, respectively. Other free parameters, including the central wavelength and amplitude, are not shown here. The fluxes of \OIIwa\ and \OIIwb\ are dependent parameters.}
\label{fig:corner_plots}
\end{figure*}

Figure~\ref{fig:corner_plots} shows the corner plots showing the posterior probability distributions from the MCMC fitting of the \OIIdw\ doublet of JD1 (top left), JD2 (top right), and the stacked spectrum (bottom). Only some parameters are shown in the corner plots, including the individual line fluxes of \OIIdw\ doublet, FWHM, and electron density. Among those parameters, only FWHM is a free parameter, while the others are dependent parameters. Other free parameters, including the peak wavelength and amplitude, are not shown in the corner plots. 

\section{Measured emission lines from the NIRSpec prism spectra}
\label{sec:lines_nirspec_prism}

The NIRSpec prism spectra of JD1 and JD2 were obtained by JWST program GO 1433 (PI Coe) and presented in \citet{Hsiao2023b}. There were two separate NIRSpec multi-object spectroscopy (MOS) observations (Obs 21 and Obs 23) using the low-resolution ($R\sim 30-300$) prism spanning a wavelength coverage of $0.6 - 5.3$ $\mu$m. Obs 23 was performed with standard 3-slitlet nods, while Obs 21 was performed with single slitlets. Each observation had a total exposure time of 1.8 hours (3.6 hours). In \citet{Hsiao2023b}, the emission lines were measured from the stacked spectrum of the four spectra (JD1 Obs 21, JD1 Obs 23, JD2 Obs 21, and JD2 Obs 23). Here, we revisit the analysis and measure the emission lines in the individual spectra. Moreover, we use the latest reduced spectra from The Dawn JWST Archive (DJA)\footnote{\url{https://dawn-cph.github.io/dja/}}, which were reduced using the upgraded \msaexp\footnote{\url{https://github.com/gbrammer/msaexp}} version 0.6.7. The NIRSpec data reduction is described in \citet{Heintz2023b} and \citet{Heintz2024}. We fit the NIRSpec prism spectra and measure the emission line fluxes using \piXedfit. Besides the individual spectra, we also fit the stacked sum of the four spectra (i.e.,~JD1 Obs 21 + JD1 Obs 23 + JD2 Obs 21 + JD2 Obs 23).

\begin{figure*}[ht]
\centering
\includegraphics[width=0.3\textwidth]{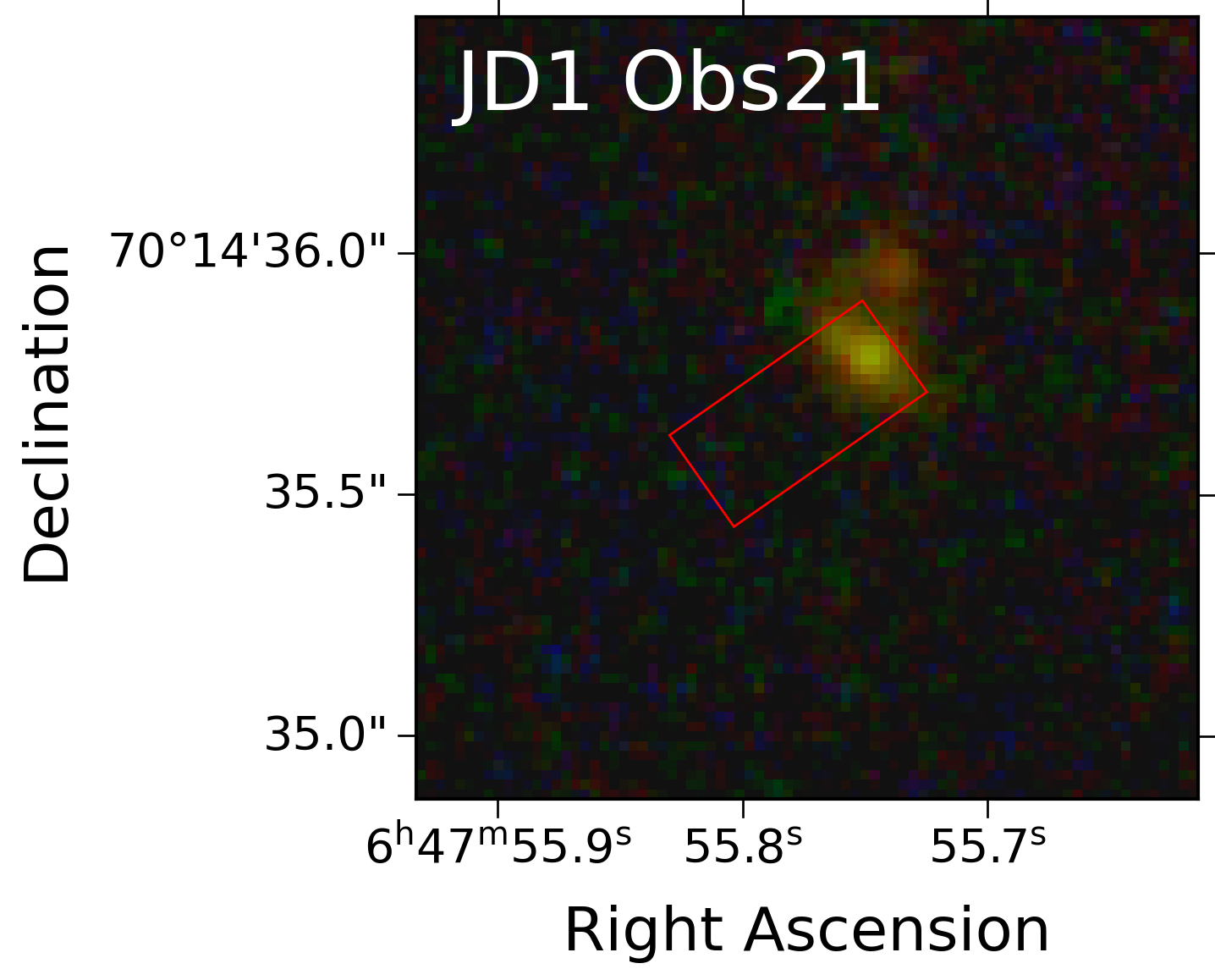}
\includegraphics[width=0.6\textwidth]{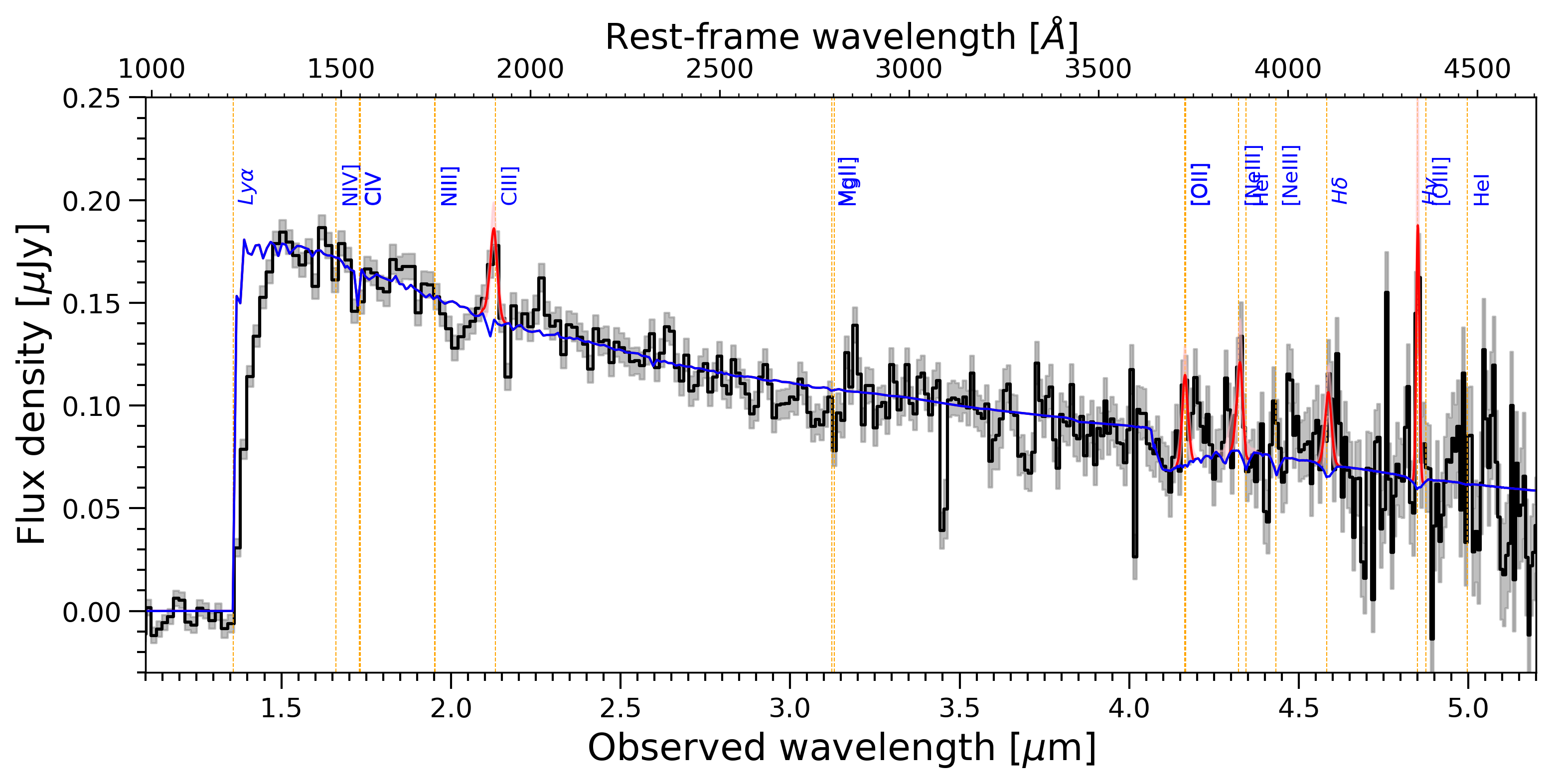}
\includegraphics[width=0.3\textwidth]{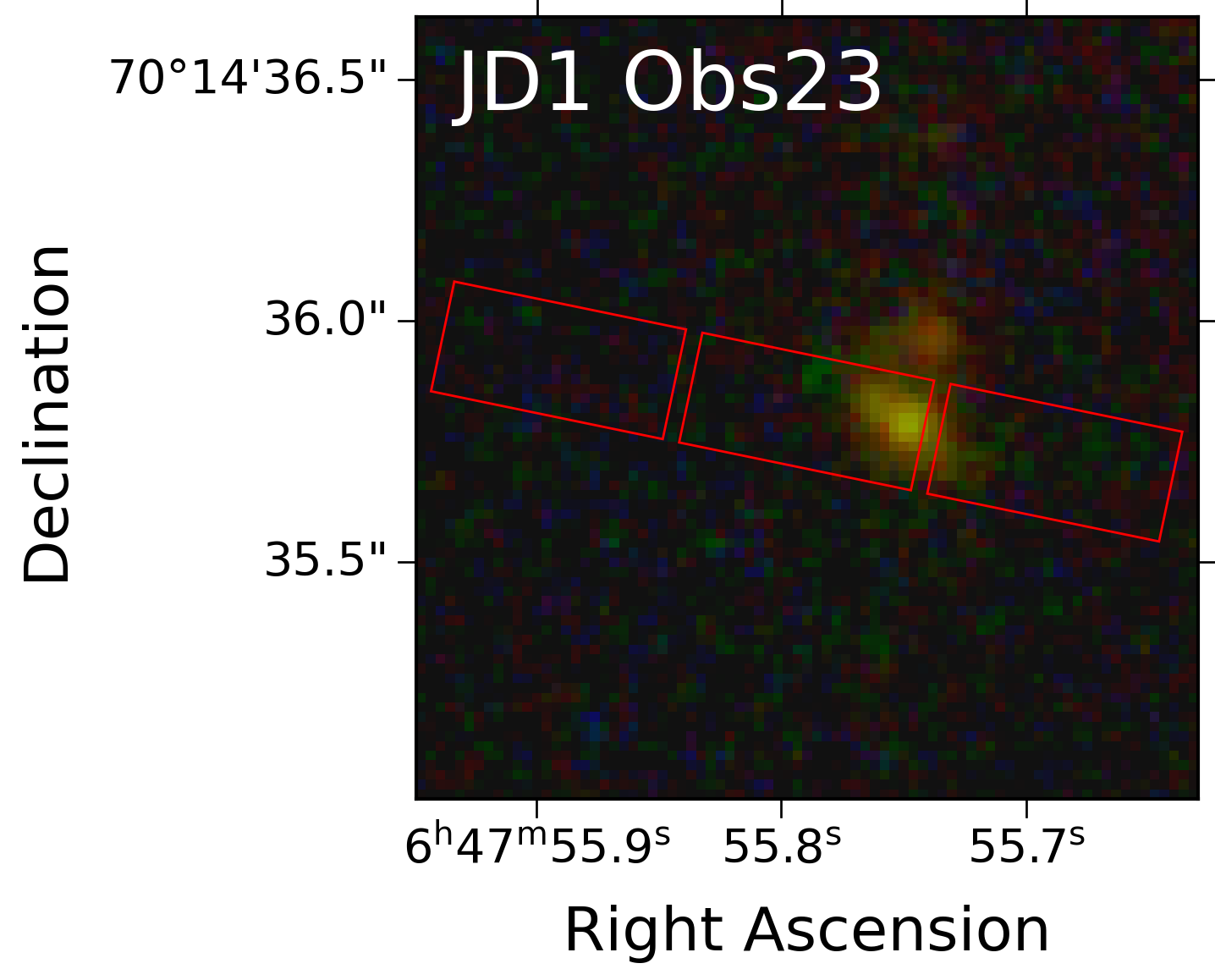}
\includegraphics[width=0.6\textwidth]{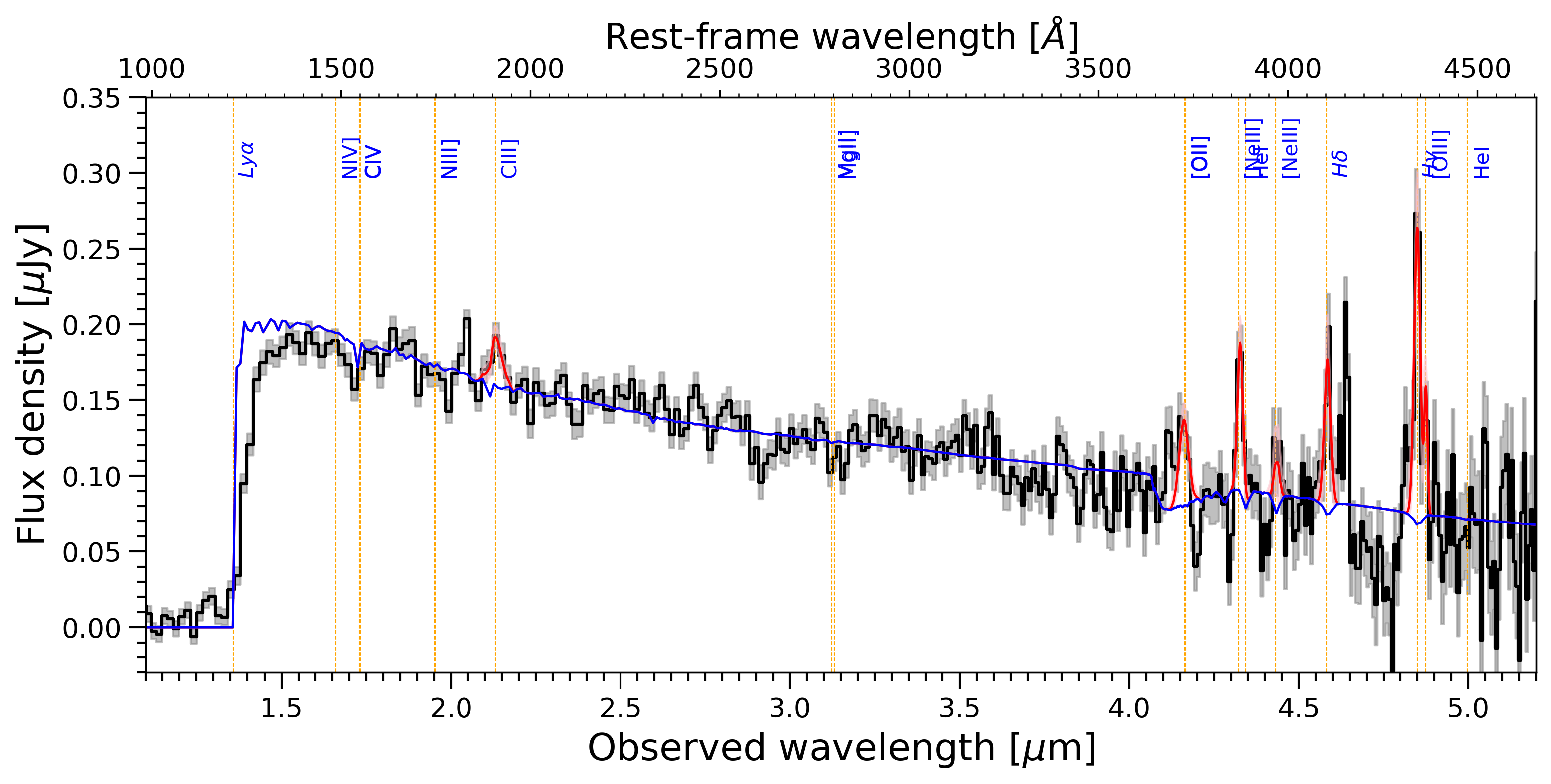}
\includegraphics[width=0.3\textwidth]{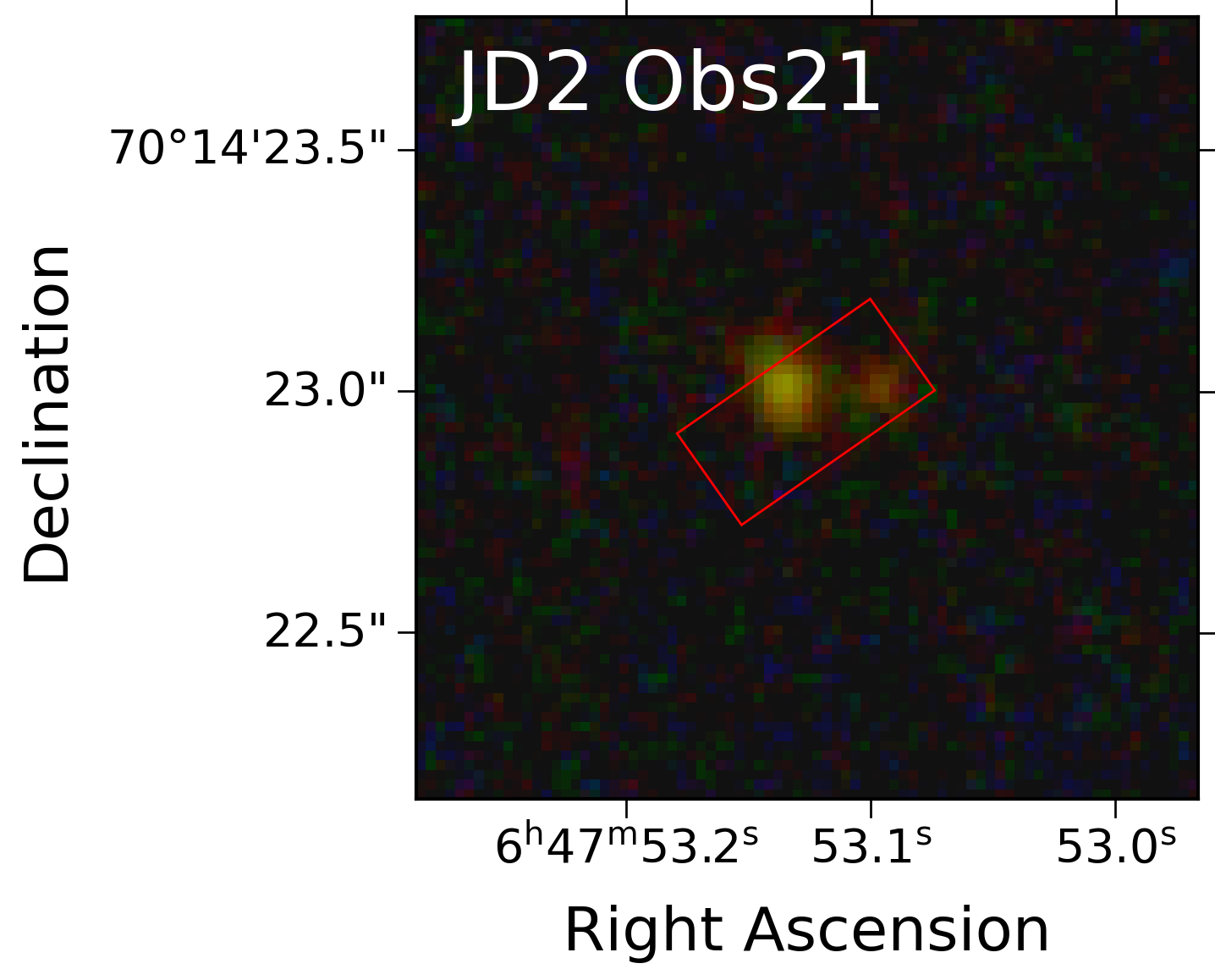}
\includegraphics[width=0.6\textwidth]{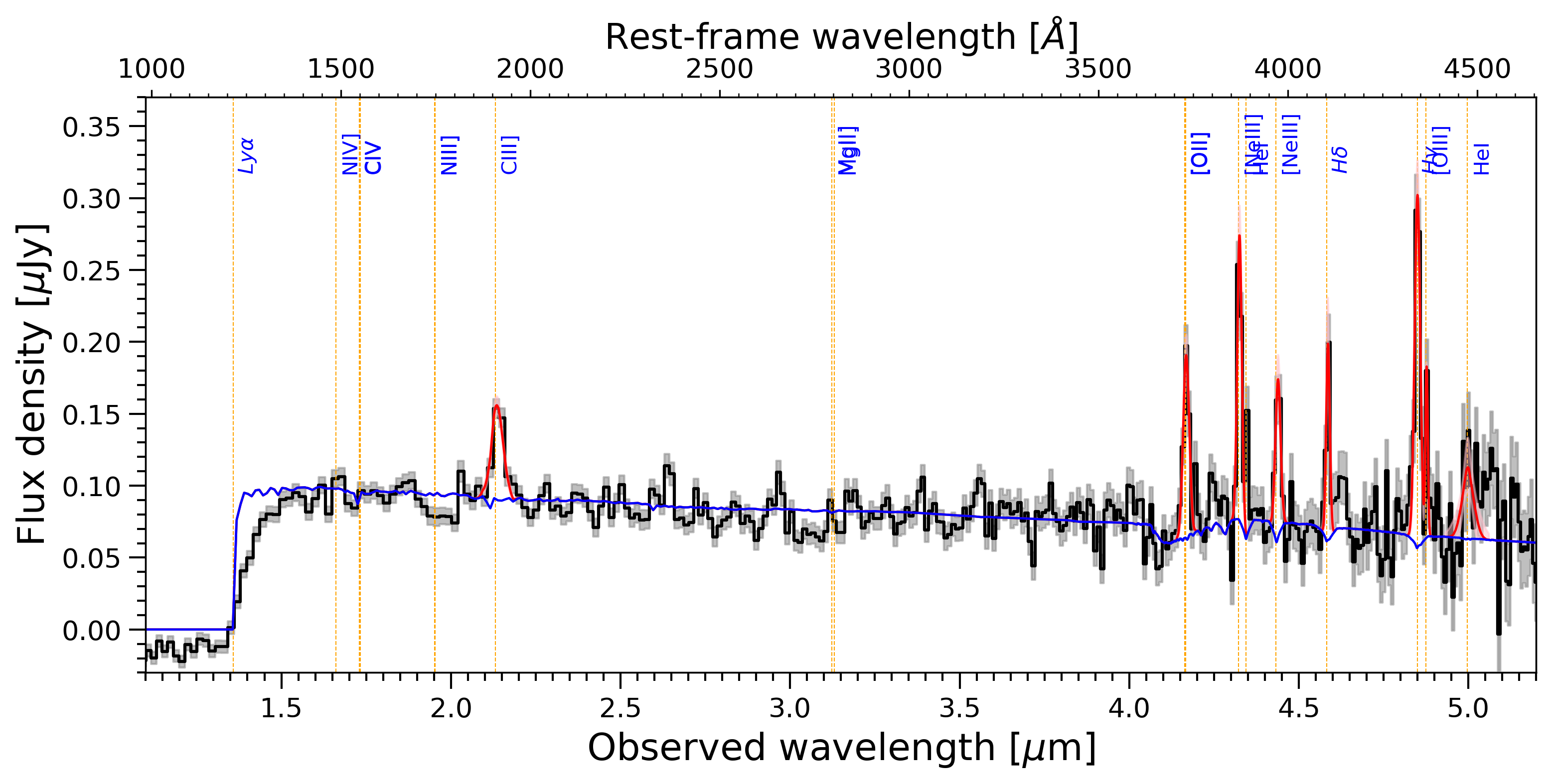}
\includegraphics[width=0.3\textwidth]{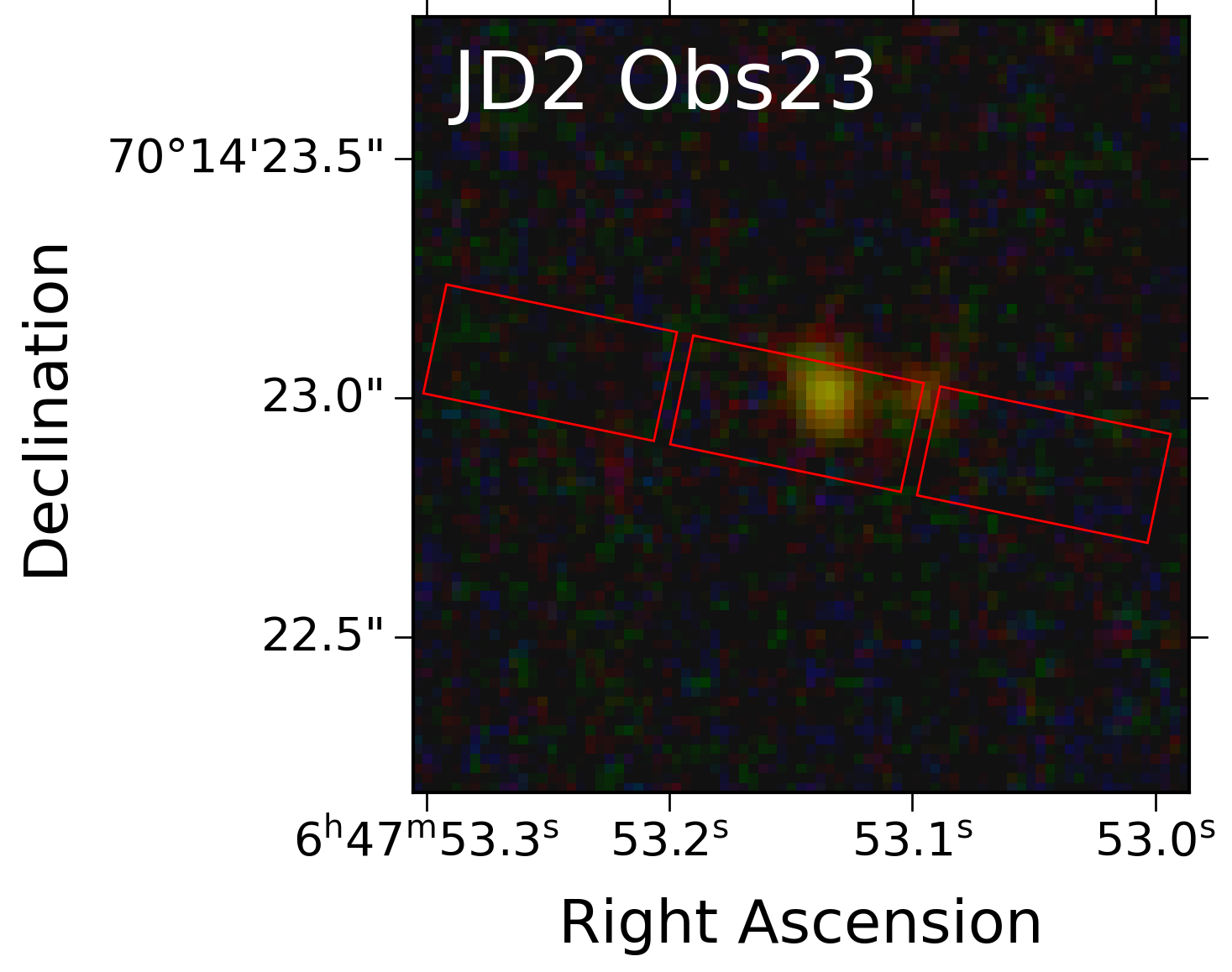}
\includegraphics[width=0.6\textwidth]{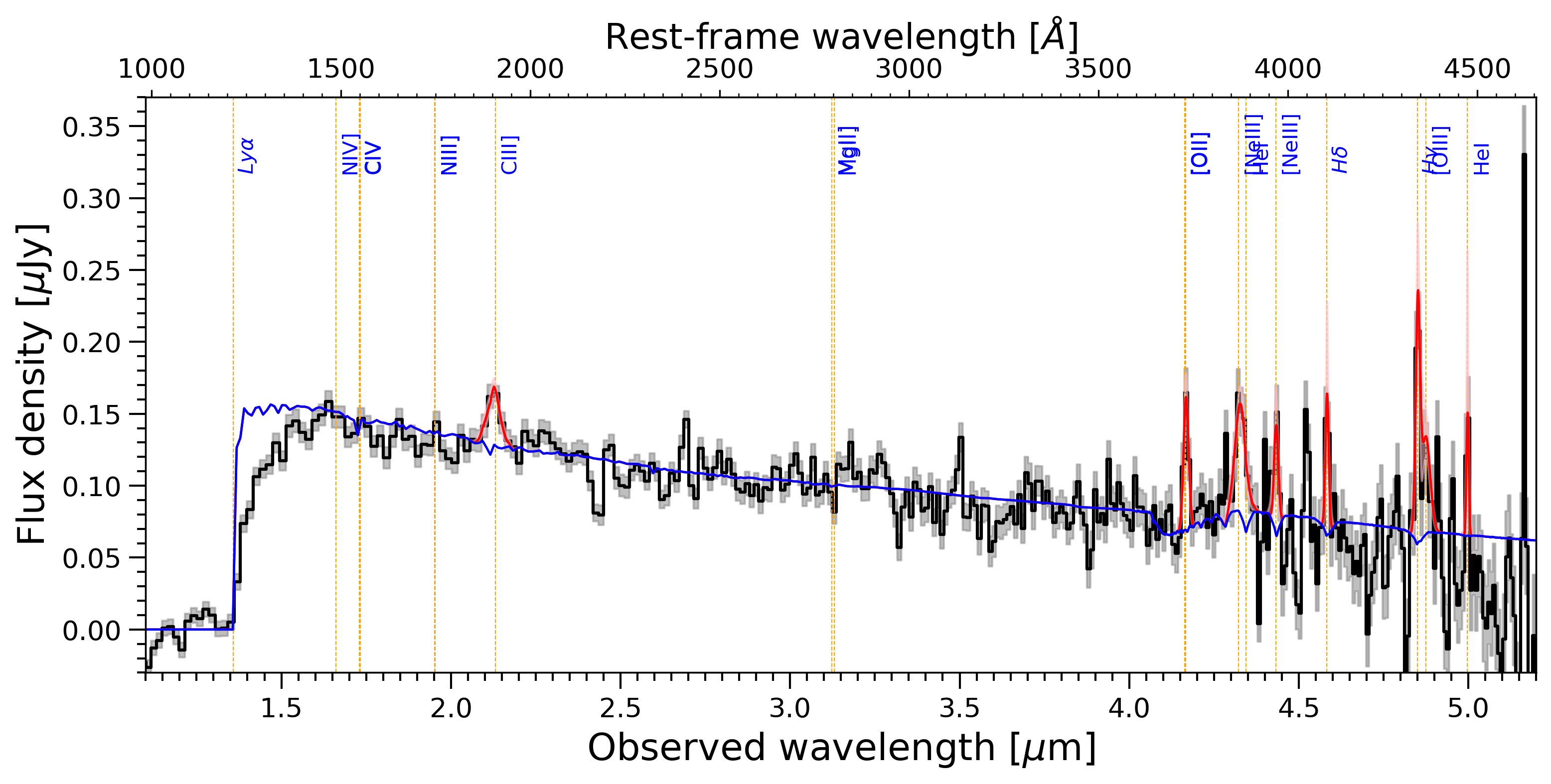}
\caption{NIRSpec prism spectra of JD1 and JD2 from two separate observations (Obs 21 and Obs 23). The observations were performed by the JWST program GO 1433 (PI Coe) and the original data were presented in \citet{Hsiao2023b}. \textit{Left}: Slitlet configurations overlaid on $1.6'' \times 1.6''$ NIRCam color images. Note that most of the observations target the brighter component A. Only JD2 Obs 21 covers both components A and B. \textit{Right}: NIRSpec prism spectra (black line and gray shaded area) and the best-fit model continuum (blue) and Gaussian for the emission lines (red). The SED modeling does not include the Lyman-$\alpha$ damping wing which makes the discrepancy between the best-fit continuum and the observed spectrum around the Lyman-$\alpha$ wavelength.}
\label{fig:nirspec_prism}
\end{figure*}

\begin{figure*}
\centering
\includegraphics[width=0.75\textwidth]{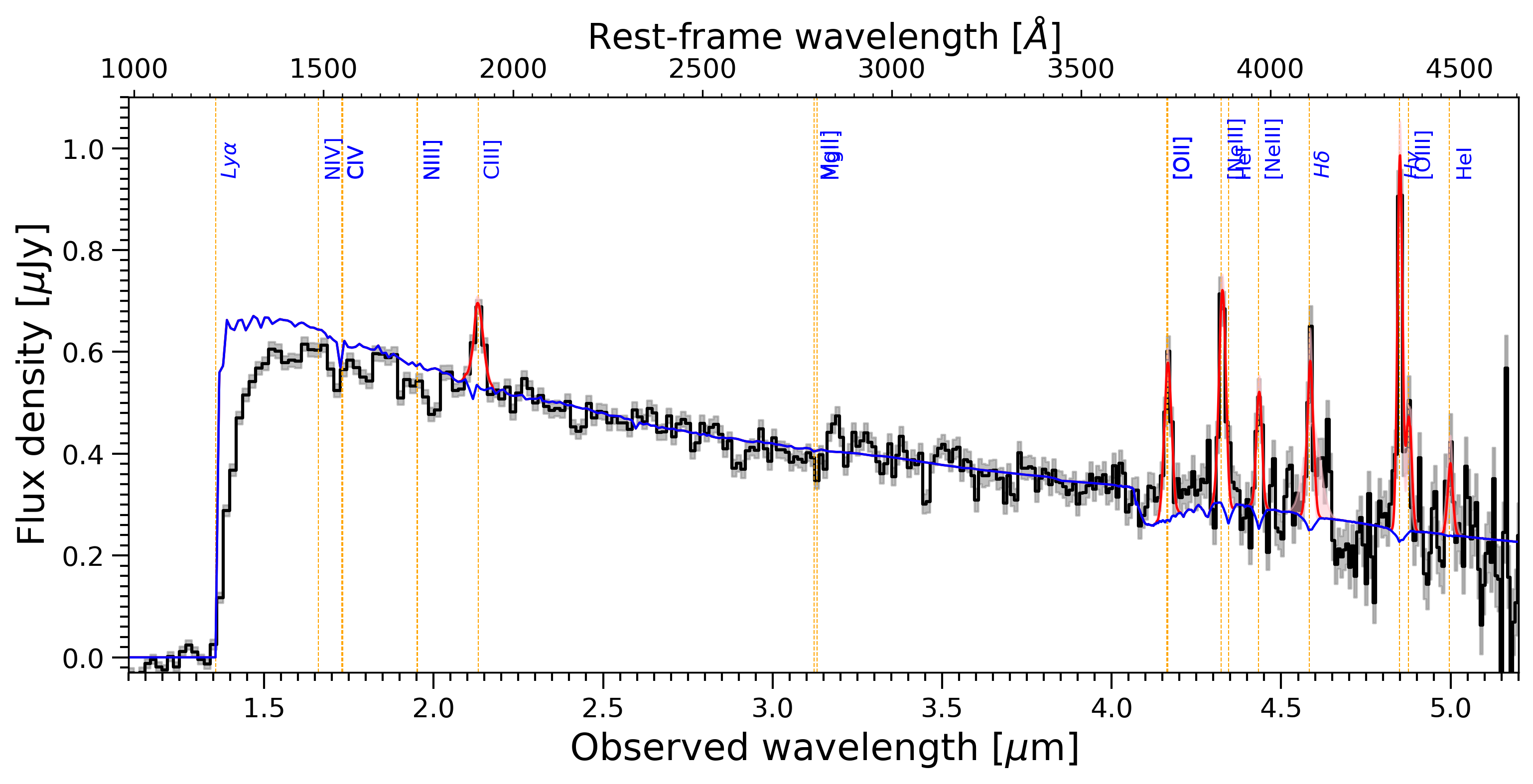}
\caption{The stacked sum of the four spectra shown in Figure~\ref{fig:nirspec_prism} (i.e.,~JD1 Obs 21 + JD1 Obs 23 + JD2 Obs 21 + JD2 Obs 23) and its fitting result. The line and symbols are the same as those in Figure~\ref{fig:nirspec_prism}.}
\label{fig:nirspec_prism_stacked}
\end{figure*}

The fitting results for the individual spectra are shown in Figure~\ref{fig:nirspec_prism} alongside the slitlets configurations overlaid on $1.6'' \times 1.6''$ NIRCam color images. We show the fitting result of the stacked spectrum in Figure~\ref{fig:nirspec_prism_stacked}. We note that most of the observations target the brighter component A. Only Obs 21 on JD2 covers both components A and B. We also note that slitlets do not cover the same area in the two observations, causing the differences in emission line fluxes. The black line and gray shaded area represent the observed spectra and their 1$\sigma$ uncertainties. The blue and red lines represent the best-fit continuum and Gaussian models of the detected emission lines. The spectra roll off gradually from $\sim 1.5$~$\mu$m (1343~\AA rest-frame) to the Lyman break observed at $1.36$~$\mu$m (1216~\AA), instead of a sharp break. This is due to the Lyman-$\alpha$ damping wing \citep[see][]{Hsiao2023b,Heintz2023b}. The continuum modeled here does not include the modeling of Lyman-$\alpha$ damping wing which makes the discrepancy between the models and observed spectra around the wavelength range. The best-fit continuum is obtained from SED fitting on the spectrum after masking the emission lines and cutting at $1.5<\lambda<5.2$ $\mu$m.   

Table~\ref{tab:lines_prism} summarizes the measured emission line properties. The line fluxes are not corrected for the lensing magnifications ($8.0$ and $5.3$ for JD1 and JD2, respectively) and slit loss. The slit loss corrections can be estimated by taking the ratio between the total aperture photometry in NIRCam filters, obtained from the \grizli\ pipeline \citep{Brammer2022} \citep[see][]{Hsiao2023b}, and pseudo photometry obtained by integrating the NIRSpec spectra through the NIRCam filters. Then the correction factors at the emission lines can be estimated by interpolating the wavelength-dependent flux ratio. The correction factors for the listed lines in Table~\ref{tab:lines_prism} are the following. JD1 Obs 21 = [2.5, 3.7, 3.9, 4.0, 4.2, 4.6, 4.6, 4.6], JD1 Obs 23 = [2.2, 3.3, 3.4, 3.5, 3.7, 4.0, 4.0, 4.0], JD2 Obs 21 = [2.7, 2.4, 2.4, 2.4, 2.3, 2.0, 2.0, 2.0], and JD2 Obs 23 = [2.0, 2.5, 2.6, 2.6, 2.7, 2.8, 2.8, 2.8].      
                   
\begin{deluxetable*}{lccccc}
\tablecaption{\label{tab:lines_jd1}Measured Emission Line Properties of the JD1 Spectrum
}
\tablewidth{\columnwidth}
\tablehead{
\colhead{Emission Line} &
\colhead{Rest wavelength} &
\colhead{Observed wavelength} &
\colhead{Observed Flux$^{a}$} &
\colhead{S/N} &
\colhead{Observed FWHM$^{b}$}
\vspace{-0.07in}\\
\colhead{} &
\colhead{(\AA)} &
\colhead{($\mu$m)} &
\colhead{($10^{-20}$ \cgsfluxunits)} &
\colhead{} &
\colhead{(\kms)}
}
\startdata
\OII
& 3727.10
& $4.16086_{-0.00034}^{+0.00030}$
& $34_{-10}^{+9}$ 
& $3.3$ 
& $175_{-22}^{+24}$\\
\OII
& 3729.86
& $4.16517_{-0.00034}^{+0.00030}$
& $23_{-9}^{+10}$ 
& $2.4$ 
& $175_{-22}^{+24}$\\
\NeIII
& 3869.86
& $4.32154_{-0.00011}^{+0.00010}$ 
& $121_{-11}^{+11}$
& $11.2$
& $180_{-22}^{+25}$\\
\HeI
& 3889.75 
& $4.34390_{-0.00051}^{+0.00046}$
& $41_{-11}^{+11}$ 
& $3.8$ 
& $184_{-22}^{+23}$\\
\NeIII
& 3968.59 
& $4.43389_{-0.00045}^{+0.00048}$
& $39_{-9}^{+9}$ 
& $3.4$ 
& $188_{-23}^{+25}$\\
\Hdelta
& 4102.89 
& $4.58160_{-0.00038}^{+0.00073}$ 
& $34_{-17}^{+11}$
& $2.2$ 
& $162_{-15}^{+20}$ \\
\Hgamma
& 4341.69
& $4.84823_{-0.00018}^{+0.00015}$
& $63_{-9}^{+9}$
& $7.0$
& $152_{-18}^{+18}$\\
\OIII
& 4364.44 
& $4.87314_{-0.00070}^{+0.00095}$
& $37_{-11}^{+11}$ 
& $3.4$ 
& $167_{-21}^{+20}$\\
\HeI
& 4472.74 
& $4.99419_{-0.00138}^{+0.00229}$
& $26_{-19}^{+20}$ 
& $1.5$ 
& $152_{-20}^{+22}$
\enddata
\tablenotetext{a}{Measured fluxes from the observed spectrum that have not been corrected for a lensing magnification, which is $\mu=8$ for JD1.}
\tablenotetext{b}{Observed FWHM before correcting for the instrumental broadening.}
\end{deluxetable*}
\begin{deluxetable*}{lccccc}
\tablecaption{\label{tab:lines_jd2}Measured Emission Line Properties of the JD2 Spectrum
}
\tablewidth{\columnwidth}
\tablehead{
\colhead{Emission Line} &
\colhead{Rest wavelength} &
\colhead{Observed wavelength} &
\colhead{Observed Flux$^{a}$} &
\colhead{S/N} &
\colhead{Observed FWHM$^{b}$}
\vspace{-0.07in}\\
\colhead{} &
\colhead{(\AA)} &
\colhead{($\mu$m)} &
\colhead{($10^{-20}$ \cgsfluxunits)} &
\colhead{} &
\colhead{(\kms)}
}
\startdata
\OII
& 3727.10
& $4.16192_{-0.00035}^{+0.00032}$
& $22_{-8}^{+8}$ 
& $2.8$ 
& $156_{-31}^{+37}$\\
\OII
& 3729.86
& $4.16500_{-0.00035}^{+0.00032}$
& $29_{-8}^{+8}$ 
& $3.4$ 
& $156_{-31}^{+37}$\\
\NeIII
& 3869.86
& $4.32203_{-0.00017}^{+0.00016}$ 
& $37_{-6}^{+6}$
& $6.0$
& $143_{-25}^{+33}$\\
\HeI
& 3889.75 
& $4.34394_{-0.00327}^{+0.00172}$
& $7_{-5}^{+7}$ 
& $1.2$ 
& $130_{-50}^{+47}$\\
\NeIII
& 3968.59 
& $4.43354_{-0.00246}^{+0.00064}$
& $11_{-6}^{+5}$ 
& $1.9$ 
& $149_{-46}^{+44}$\\
\Hdelta
& 4102.89 
& $4.58226_{-0.00091}^{+0.00074}$ 
& $12_{-9}^{+8}$
& $1.6$ 
& $125_{-32}^{+42}$ \\
\Hgamma
& 4341.69
& $4.84911_{-0.00057}^{+0.00082}$
& $30_{-12}^{+12}$
& $2.5$
& $144_{-43}^{+39}$\\
\OIII
& 4364.44 
& $4.87172_{-0.00149}^{+0.00278}$
& $6_{-4}^{+4}$ 
& $1.4$ 
& $102_{-39}^{+44}$\\
\HeI
& 4472.74 
& $4.34394_{-0.00327}^{+0.00172}$
& $7_{-5}^{+7}$ 
& $1.2$ 
& $130_{-50}^{+47}$
\enddata
\tablenotetext{a}{Measured fluxes from the observed spectrum that have not been corrected for a lensing magnification, which is $\mu=5.3$ for JD2.}
\tablenotetext{b}{Observed FWHM before correcting for the instrumental broadening.}
\end{deluxetable*}

\begin{deluxetable*}{ccccc}
\tablecaption{\label{tab:poly_coeff}Coefficients for the polynomial function in Eq.~\ref{eq:poly_function}}
\tablewidth{\columnwidth}
\tablehead{
\colhead{Coefficient} &
\colhead{$a$} &
\colhead{$b$} &
\colhead{$c$} &
\colhead{$d$}
}
\startdata
$\alpha$
& $-0.00447$
& $0.28813$
& $-0.06841$ 
& $0.00506$ \\
$\beta$
& $-7616.41042$
& $7075.85077$
& $-2433.11054$ 
& $321.56781$ \\
$\gamma$
& $-1661.48935$
& $1481.10495$
& $-507.35920$ 
& $69.67036$
\enddata
\end{deluxetable*}

\begin{deluxetable*}{lcccccc}
\tablecaption{\label{tab:lines_prism}Measured Emission Line Properties from NIRSpec prism spectra
}
\tablewidth{\columnwidth}
\tablehead{
\colhead{Emission Line} &
\colhead{$\lambda$ rest} &
\colhead{JD1 Obs 21 Flux} &
\colhead{JD1 Obs 23 Flux} &
\colhead{JD2 Obs 21 Flux} &
\colhead{JD2 Obs 23 Flux} &
\colhead{Stacked Flux}
\vspace{-0.07in}\\
\colhead{(1)} &
\colhead{(2)}  &
\colhead{(3)} &
\colhead{(4)} &
\colhead{(5)} &
\colhead{(6)} &
\colhead{(7)}
}
\startdata
\CIII
& 1906.68,1908.73
& $95_{-14}^{+15}$
& $98_{-21}^{+21}$ 
& $194_{-17}^{+17}$
& $129_{-20}^{+21}$
& $428_{-35}^{+34}$ \\
\OII
& 3727.10,3729.86
& $17_{-4}^{+5}$
& $34_{-7}^{+7}$ 
& $50_{-5}^{+5}$ 
& $30_{-5}^{+5}$
& $144_{-12}^{+12}$ \\
\NeIII
& 3869.86
& $19_{-5}^{+6}$
& $35_{-5}^{+6}$ 
& $54_{-5}^{+5}$ 
& $51_{-7}^{+8}$
& $177_{-12}^{+12}$ \\
\NeIII
& 3968.59
& \nodata
& $14_{-6}^{+8}$ 
& $33_{-4}^{+5}$ 
& $20_{-5}^{+5}$
& $88_{-10}^{+10}$ \\
\Hdelta
& 4102.89
& $15_{-5}^{+5}$
& $34_{-7}^{+7}$ 
& $27_{-4}^{+5}$ 
& $19_{-5}^{+5}$
& $117_{-17}^{+55}$ \\
\Hgamma
& 4341.69
& $21_{-4}^{+4}$
& $58_{-8}^{+9}$
& $65_{-6}^{+6}$ 
& $34_{-6}^{+6}$
& $179_{-12}^{+13}$ \\
\OIII
& 4364.44 
& \nodata
& $12_{-3}^{+3}$ 
& $13_{-3}^{+3}$ 
& $28_{-7}^{+7}$
& $62_{-5}^{+5}$ \\
\HeI
& 4472.735 
& \nodata
& \nodata 
& $26_{-8}^{+9}$ 
& $12_{-6}^{+6}$
& $36_{-12}^{+13}$
\enddata
\tablecomments{(1) Emission line. (2) Rest-frame wavelength. (3) Measured line fluxes of the JD1 Obs 21 spectrum. (4) Measured line fluxes of the JD1 Obs 23 spectrum. (5) Measured line fluxes of the JD2 Obs 21 spectrum. (6) Measured line fluxes of the JD2 Obs 23 spectrum. (7) Measured line fluxes from the stacked sum of the four spectra (JD1 Obs21 + JD1 Obs23 + JD2 Obs21 + JD2 Obs23). All flux densities are in units of $10^{-20}$ \cgsfluxdensityunits. The line fluxes are not corrected for the lensing magnifications (8.0 and 5.3 for JD1 and JD2, respectively) and slit loss.}
\end{deluxetable*}


\bibliography{macs0647jd}{}
\bibliographystyle{aasjournal}



\end{document}